\documentclass[epj]{svjour}
\usepackage{epsf}
\usepackage{amsmath,amsfonts,amstext,theorem,amssymb}
\begin{document}

\title{Andreev-Lifshitz supersolid revisited for a few  
electrons on a square lattice I}
\titlerunning{Andreev-Lifshitz supersolid I}

\author{Georgios Katomeris, Franck Selva \and Jean-Louis Pichard}

\authorrunning{G. Katomeris et al}
        
\institute{CEA/DSM, Service de Physique de l'\'Etat Condens\'e, 
Centre d'\'Etudes de Saclay, 91191, Gif sur Yvette cedex, France}

\abstract
{ In 1969, Andreev and Lifshitz have conjectured the existence of a 
supersolid phase taking place at zero temperature between the quantum 
liquid and the solid. In this and a succeeding paper, we re-visit this 
issue for a few polarized electrons (spinless fermions) interacting via 
a $U/r$ Coulomb repulsion on a two dimensional $L \times L$ square 
lattice with periodic boundary conditions and nearest neighbor hopping 
$t$. This paper is restricted to the magic number of particles $N=4$ 
for which a square Wigner molecule is formed when $U$ increases and 
to the size $L=6$ suitable for exact numerical diagonalizations. When 
the Coulomb energy to kinetic energy ratio $r_s=UL/(2t\sqrt{\pi N})$ 
reaches a value $r_s^F \approx 10$, there is a level crossing between 
ground states of different momenta. Above $r_s^F$, the mesoscopic 
crystallization proceeds through an intermediate regime ($r_s^F <r_s 
<r_s^W \approx 28$) where unpaired fermions with a reduced Fermi energy 
co-exist with a strongly paired, nearly solid assembly. We suggest that 
this is the mesoscopic trace of the supersolid proposed by Andreev and 
Lifshitz. When a random substrate is included, the level crossing at 
$r_s^F$ is avoided and gives rise to a lower threshold $r_s^F(W) < r_s^F$ 
where two usual approximations break down: the Wigner surmise for the 
distribution of the first energy excitation and the Hartree-Fock 
approximation for the ground state.
}


\maketitle

\section{Introduction}
\label{sec:intro}

 A basic question in quantum many body theory is to know how 
one goes from independent particle motion towards collective 
motion when one decreases the density $n_s$ of a system of 
charged particles repelling each other via a $U/r$ Coulomb 
repulsion. As introduced long ago by Wigner, the factor $r_s$, 
defined as the radius $r$ of the volume enclosing a single particle 
in units of the Bohr radius $a_B$, governs this crossover. 
Quantum mechanical effects are important when $r_s$ is small, 
and become more and more negligible when $r_s$ becomes large. 
Since $r_s \propto 1/{\sqrt n_s}$ in two dimensions, the quantum 
limit of a Fermi liquid is obtained at large densities $n_s$. In 
the dilute limit, the quantum effects disappear and the charges 
crystallize, forming a Wigner solid of minimum electrostatic 
energy. A calculation of the electrostatic energy of different 
crystalline arrays shows that the hexagonal array minimizes the 
energy in the $2d$ continuum. However, in a square lattice model 
with periodic boundary conditions (BCs) and for a sufficient filling 
factor, the symmetry of the Wigner solid is restricted and the 
formation of a square crystalline array is favored. 
 
 Before studying this quantum-classical crossover in a mesoscopic   
lattice model, let us mention that it is usually assumed 
\cite{ceperley,imada} that a single liquid-solid transition 
takes place at $r_s \approx 37$ in the continuous $2d$ thermodynamic 
limit. This single transition was obtained by fixed 
node quantum Monte Carlo calculations \cite{ceperley,candido}, 
allowing to study a few hundreds of electrons and to 
vary their number for estimating the finite size effects. 
When the spin degrees of freedom are included, the existence of an 
intermediate polarized liquid phase separating the unpolarized 
liquid and the Wigner solid is still debated \cite{varsano}. 
Though the quantum Monte Carlo calculations have the advantage 
to allow the study of a relatively large number of particles, 
they have the well-known ``sign problem''. This leads to fixed node 
approximations which are made to avoid the negative weights that would 
be generated otherwise by antisymmetric states, and gives only an upper 
bound to the exact ground state energy. In Refs. \cite{ceperley,candido} 
for instance, two nodal structures have been considered, given by two 
Slater-Jastrow wave functions adapted to describe the weak coupling Fermi 
liquid (nodal structure of a Slater determinant of plane waves) and 
the strong coupling Wigner solid (nodal structure of a Slater determinant 
of localized site orbitals). In these works, the existence of a single 
transition separating the Fermi liquid from the Wigner solid is 
a consequence of the assumed nodal  structures. Very recently, the ground 
state energy of the two-dimensional uniform electron gas has been 
calculated \cite{attaccalite} with a fixed node diffusion Monte Carlo 
method, including backflow correlations. The backflow method allows to 
partly relax the previous nodal constraints. The backflow nodes give 
smaller energies that the previous nodes, supporting the existence 
of an intermediate polarized liquid phase for $26 < r_s < 35$.  

\subsection{$2d$ metal and related issues}
\label{sec:metal}

 The motivation to re-visit nowadays charge crystallization in $2d$ Coulomb 
systems is fourfold. Firstly, it becomes possible to create $2d$ gases 
of charges in high quality field effect devices and to decrease by a gate 
the carrier density $n_s$ for obtaining a large factor $r_s$. Doped 
semi-conductors \cite{kravchenko} (Si-Mosfet, Ga-As heterostructures, 
Si-Ge quantum wells) can be now used to study how one goes in two 
dimensions from a Fermi liquid towards a Wigner crystal. Secondly, the 
direct observation of the Wigner crystal being difficult, one can 
nevertheless measure the conductance of the dilute $2d$ electron gas 
at different densities as a function of the temperature, of the bias 
voltage, of a parallel magnetic field, etc. Remarkably, those transport 
measurements, first done \cite{kravchenko} by Kravchenko et al in high 
quality Si-MOSFETs, show the existence of an unexpected metal-insulator 
transition (MIT) when the gate voltage varies and the dilute limit is 
reached. A similar MIT was later observed \cite{kravchenko} using a hole 
gas in Ga-As heterostructures and in Si-Ge quantum wells. This observation 
of a low temperature metallic behavior for typically $3 \ldots 6 < r_s < 
9 \ldots 30$ (the largest ratios characterizing the cleaner samples) raises 
the question of a possible intermediate phase, which should be neither a 
Fermi glass of localized particles (Anderson insulator), nor a correlated 
Wigner solid (pinned insulating crystal). Quite recently, magnetotransport 
measurements in low density $2d$ hole gas in GaAs quantum wells have been 
interpreted \cite{gao1,gao2} in terms of an unknown metallic phase 
coexisting with a Fermi liquid phase, for estimated values $12 < r_s < 18$. 
Local compressibility measurements by Ilani et al also point to a two-phase 
coexistence picture for intermediate values of $r_s$ in GaAs \cite{illani}. 
Thirdly, the formation of a mesoscopic Wigner molecule can be also 
nowadays studied using a quantum dot\cite{dot} with a few electrons 
or a few ions \cite{ion} trapped by electric and magnetic fields. 
Increasing the size of the trap yields \cite{yannouleas} a crossover 
from independent-particle towards collective motion. Quantum dots with 
a few electrons are among possible candidates for providing the Qubits 
of a future quantum computer \cite{kouwenhoven}. 
Lastly, an unexplained intermediate regime was numerically observed 
\cite{bwp1,bwp2,selva} in studying the persistent currents carried by 
the ground state and the low energy excitations of mesoscopic disordered 
clusters. Both experiments and numerics give unexplained low energy 
behaviors for similar intermediate values of the ratio $r_s$. 
  
\subsection{Andreev-Lifshitz supersolid}

  In this study, we consider fully polarized electrons and ignore 
possible magnetic transitions. For avoiding uncontrolled assumptions, 
we consider a system which is small enough to allow exact numerical 
diagonalization, but where the lattice effects and the finite size 
corrections are important. Those two effects have been studied 
in details in Ref. \cite{moises} for two polarized electrons. 
Our main goal is to characterize as precisely as possible the 
ground state for intermediate ratios $r_s$, in order to see if a 
small lattice model does not exhibit the mesoscopic signature of an 
intermediate phase separating the solid from the liquid, where the 
solid and the fluid would coexist. Such a vacancy-solid phase was 
indeed suggested \cite{andreev} by Andreev and Lifshitz if the zero 
point motions of certain defects become sufficient to form waves 
propagating inside the solid. This Andreev-Lifshitz supersolid was first 
proposed for three dimensional quantum solids made of He atoms. Castaing 
and Nozi\`eres have later considered \cite{castaing} such a possibility 
for spin polarized He$^3$. The statistics of the defects depend on their 
nature. For simple vacancies in the crystal, their statistics is given 
by the statistics of the particles out of which the solid is made. If the 
defects are bosons, they may form a condensate, giving rise to a superfluid 
coexisting with the solid. This supersolid phase is discussed in certain 
bosonic models \cite{batrouni}. If the defects are fermions, they may form 
a Fermi liquid \cite{dzyaloshinskii} coexisting with the solid, 
such that the system is neither a solid, nor a liquid. Two kinds of 
motion are possible in it; one possesses the properties of motion in 
an elastic solid, the second possesses the properties of motion in a 
liquid. 

If one considers the quantum melting of the solid from the dilute limit 
(large $r_s$),  the nature of the relevant defects is not 
an easy question. One can imagine a particle being put into an 
intersticial site of the Wigner lattice, creating a vacancy-intersticial pair 
at a certain electrostatic cost $\delta U$.  A model assuming such defects 
has been recently proposed \cite{shapiro} for describing addition spectra in 
quantum dots. Classically, this vacancy-intersticial pair remains 
localized. But quantum tunneling may lead to delocalization of the defects  
and to the appearance of a band of defects of finite width $B_d$ which 
increases when $r_s$ decreases. When $B_d$ exceeds $\delta U$, one can 
imagine two possibilities: either the total quantum melting \cite{candido} 
of the Wigner crystal (simple solid-liquid transition), or a partial melting 
leading to the persistence of a floppy crystal with delocalized defects. 
If a delocalized defect appears in the quantum crystal, the crystal remains 
perfectly periodic, but the number of crystal lattice sites becomes smaller 
than the total number of particles. This is the supersolid scenario 
proposed by Andreev-Lifshitz for He physics. 

If one considers charge crystallization from the other limit, where 
the density $n_s$ is large (small $r_s$),  
one can argue that the interaction will create correlated pairs of 
particles near the Fermi surface, but will not reorganize the one particle 
states well below the Fermi surface. Such a possibility has been proposed   
by Bouchaud et al \cite{bouchaud} for liquid He$^3$. Moreover, they have 
developed a variational approach, based on a fixed number of fermions BCS 
wave function, having a different nodal structure than the Jastrow-Slater 
nodal structures considered in Ref. \cite{ceperley}. In this picture, the 
system is thought as made of unpaired fermions with a {\it reduced} Fermi 
energy, co-existing with a strongly paired, nearly solid assembly. 
Furthermore, it was stressed that the supersolid - first 
introduced by Andreev and Lifshitz - would be a good candidate to describe 
this new phase. To the concept of a crystal with a reduced number of 
crystal lattice sites, as discussed by Andreev and Lifshitz 
from the solid limit, corresponds the concept of unpaired fermions with 
a reduced Fermi energy, as discussed by Bouchaud et al from the 
liquid limit.
  
 Our purpose is to study if a supersolid regime is relevant for 
describing a low density $2d$ electron (or hole) gas. This numerical work 
being restricted to very small system sizes, the study is limited to 
merely study if one can detect the mesoscopic trace of a possible 
supersolid. One of our motivations comes from the observation \cite{bwp2} 
that the low energy levels do not obey Wigner-Dyson statistics for 
disordered clusters at intermediate ratios $r_s$. This will be again 
emphasized at the end of this work for the statistics of the first energy 
excitation. This suggests the existence of low energy collective 
excitations in the clean limit. For this reason, we have studied 
clusters without random substrate and we have observed at intermediate 
$r_s$ a floppy correlated solid coexisting with a liquid of unpaired 
particles. This conclusion is supported by a study of the projection 
of the ground state (GS) onto a combination of Slater 
determinants (SDs) built out from plane waves and from site orbitals. 
The plane wave SDs are given by the low energy levels of same total 
momentum $\vec{K}$ as the intermediate GS, and correspond to unpaired 
fermions with a reduced Fermi energy. The site SDs describe the Wigner 
solid molecule and its small fluctuations. Since the GS is given by the 
combination of unpaired fermions and of a floppy Wigner molecule for 
$r_s^F \approx 9.3 < r_s <r_s^W \approx 28$ in the studied system, we 
suggest that this is the mesoscopic trace of the supersolid 
discussed in Refs. \cite{andreev} and \cite{bouchaud}. The study of the 
GS response to various perturbations (Aharonov-Bohm flux, pinning well) 
and of the distributions of the different inter-particle spacings allows 
us to give a few remarkable properties of the intermediate 
regime. Eventually, we consider the effect of disorder and give 
further evidence of the existence of a first threshold $r_s^F(W)$ where 
strong correlation effects occur without yielding a full crystallization: 
(i) the breakdown of the Hartree-Fock approximation for the ground state 
and (ii) the breakdown of Wigner-Dyson level repulsion for the first 
excitation. 

\section{Lattice model}
\label{sec:Hamiltonian}

We consider fully polarized electrons (i.e. spinless fermions), 
having symmetric spin wave functions and antisymmetric orbital wave 
functions, free to move in an $L \times L$ lattice with periodic BCs, 
and interacting via a $U/\vec{r}$ Coulomb repulsion. The Hamiltonian 
reads
\begin{equation}
H = -t \sum_{<\vec{i},\vec{j}>} c_{\vec{i}}^{\dagger} c_{\vec{j}} 
+ \sum_{\vec{i}} v_{\vec{i}} n_{\vec{i}} + 
\frac{U}{2} \sum_{{\vec{i}}\neq {\vec{j}}} \frac{n_{\vec{i}} n_{\vec{j}}}
{|\vec{r}_{\vec{ij}}|}
\label{siteH}
\end{equation}
where $\vec{i}$, $\vec{j}$ label the lattice sites, 
$\left<\vec{i},\vec{j}\right>$ 
means $\vec{i}$ nearest neighbor to $\vec{j}$, $c_{\vec{i}}^{\dagger}$, 
$c_{\vec{i}}$ are the creation, annihilation 
operators of a spinless fermion at the site $\vec{i}$; $n_{\vec{i}}=
c_{\vec{i}}^{\dagger}c_{\vec{i}}$ 
is the occupation number at the site labeled by the vector 
$\vec{i}=(i_x,i_y)$. The vector $\vec{r_{ij}}$ is 
defined as the shortest vector going from the site $\vec{i}$ to the site 
$\vec{j}$ in a square lattice with periodic BCs ($r_{x}$ and $r_y \leq L/2$). 
$t=\hbar^2/(2m a^2)$ is the hopping term, $a$ the lattice spacing, 
$v_{\vec{i}}$ the site potentials which are randomly distributed 
in the interval $[-W/2, W/2]$ and $U=e^2/(\epsilon a)$ the Coulomb 
interaction between two fermions separated by $a$ in a medium of 
dielectric constant $\epsilon$. 

This work is restricted to a detailed study of the case $N=4$ and $L=6$, 
corresponding to a filling factor $\nu=N/L^2=1/9$. $N=4$ is a 
`magic' number for which at large values of $U$, the $3 \times 3$ 
square Wigner molecule is commensurate with the imposed $6 \times 6$ square 
lattice. The $r_s$ factor, defined in the continuum as 
\begin{equation}
r_s=\frac{1}{\sqrt{\pi n_s} a_B}
\end{equation}
for a carrier density $n_s$ and a Bohr radius $a_B=\hbar^2\epsilon/(me^2)$, 
becomes in a lattice model
\begin{equation}
r_s=\frac{U}{2t\sqrt{\pi \nu}}
\end{equation} 
since $\hbar^2/(2m a^2) \rightarrow t$, $e^2/(\epsilon a) \rightarrow U$. 
and $n_s=\nu/a^2$.
 
 A $L \times L$ continuous $2d$ torus having infinitely more 
degrees of freedom than a mere $6 \times 6$ lattice, one cannot compare 
the obtained lattice behaviors to those obtained assuming a continuous 
space, as in Ref. \cite{ceperley}, without further investigations. 
Nevertheless, appropriately defined observables should only depend 
on the value of the dimensionless ratio $r_s$, up to certain finite size 
corrections. This has been checked \cite{moises} for two polarized 
electrons on a square lattice, when the fluctuation $\Delta r$ of the 
distance $r$ between the two particles is larger than the lattice spacing 
$a$. Calculating $\Delta r$ in powers of $t/U$, one finds \cite{moises} 
that a correlated lattice regime takes place when $r_s$ exceeds a 
threshold value $r_s^* \approx 100$ when $N=2$ and $L=6$. Below $r_s^*$, 
$r_s \propto UL/t$ is the relevant scaling variable, up to certain finite 
size corrections of order $1/L^2$. Above $r_s^*$, one has a lattice regime 
where $r_s$ is not a relevant scaling variable. If the threshold 
value $r_s^*$ does not vary very much when one goes from $N=2$ to 
$N=4$, one has a chance to observe a four particle Wigner molecule free of 
important lattice effects as far as $r_s < r_s^* \approx 100$ when 
$L=6$. 

\section{The non disordered lattice}
\label{sec:clean}

\begin{figure}
\centerline
{ 
\epsfxsize=4.2cm 
\epsffile{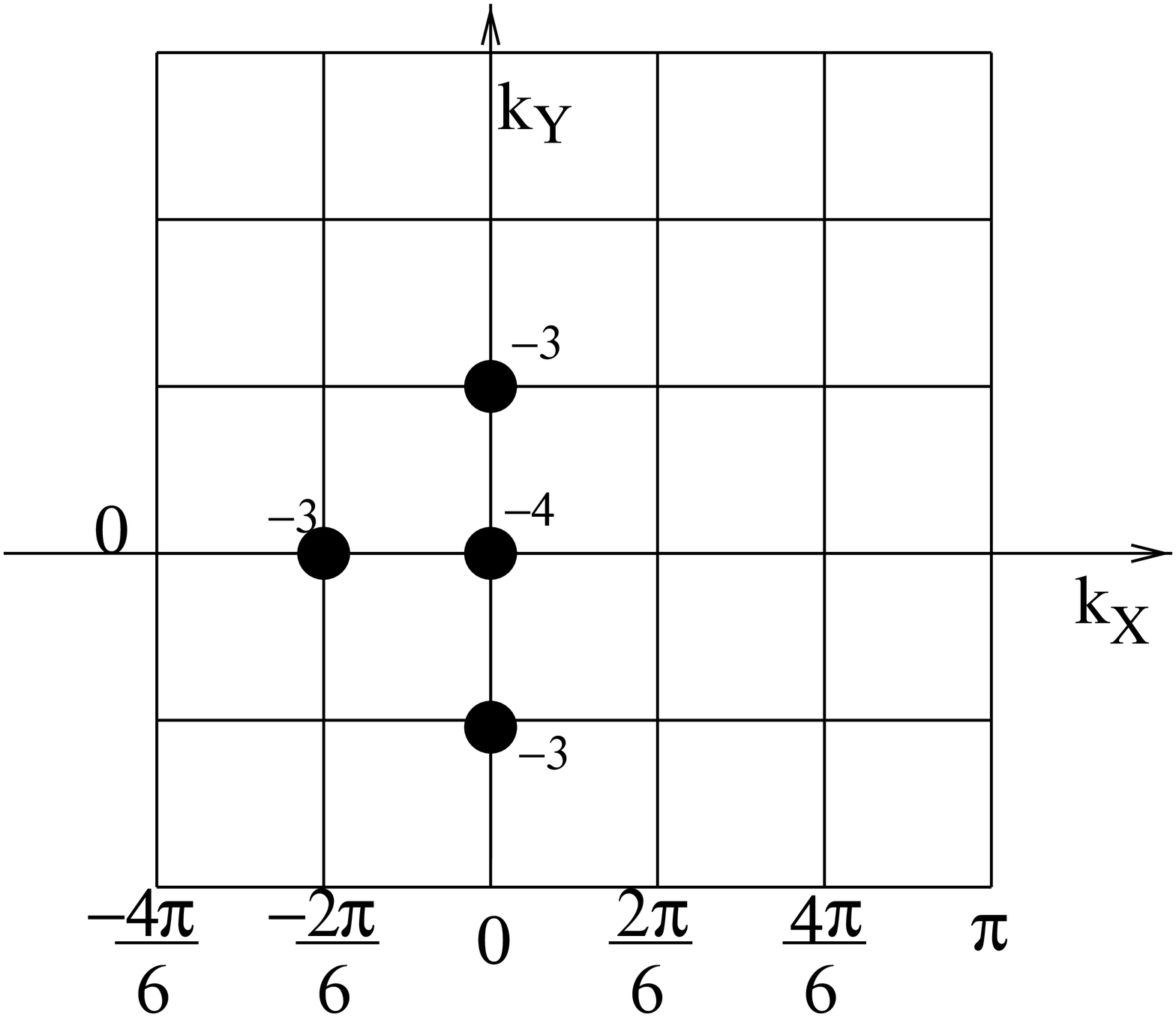}
\hfill 
\epsfxsize=4.2cm  
\epsffile{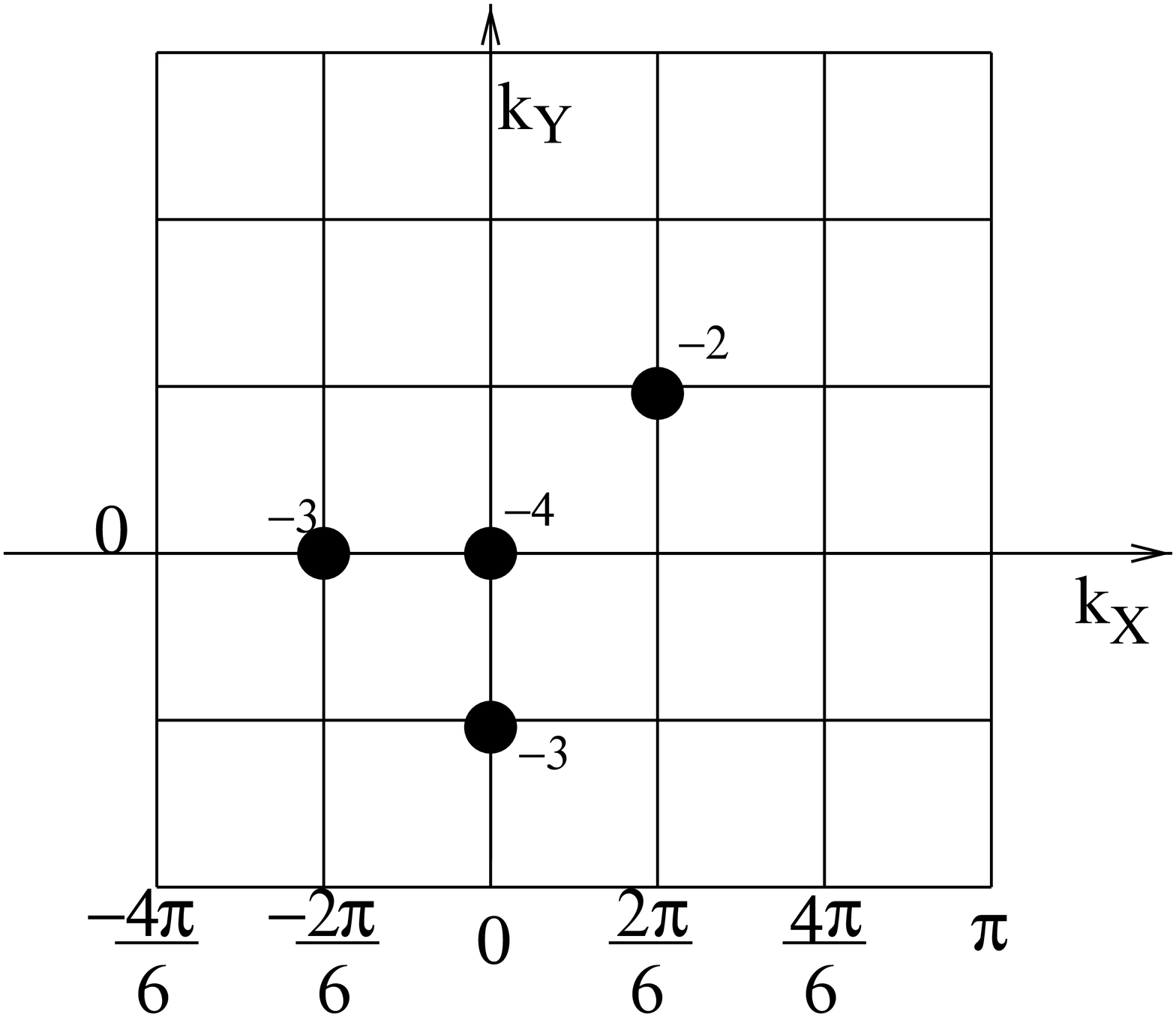}
}
\centerline
{
\epsfxsize=4.2cm  
\epsffile{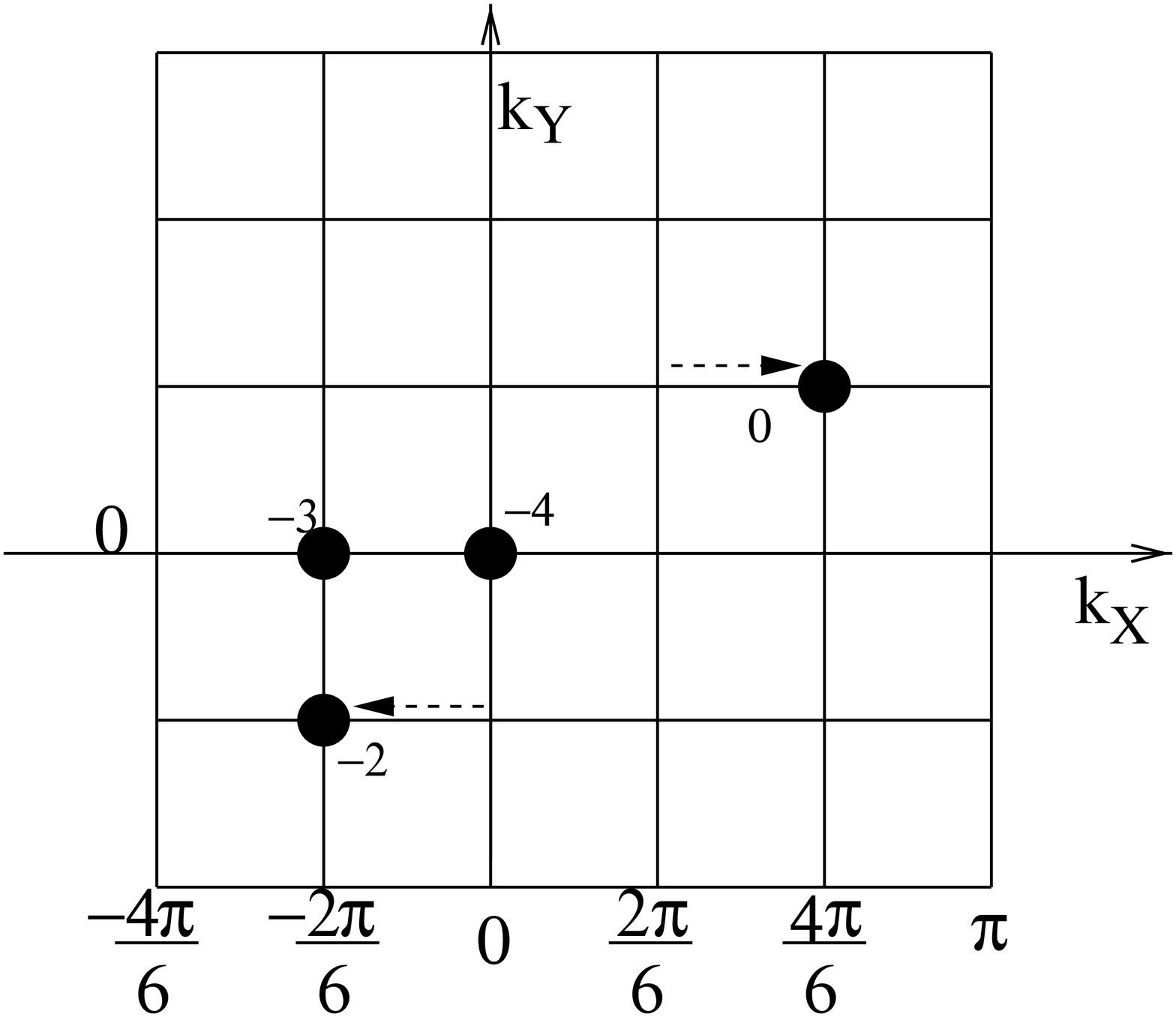}
\hfill 
\epsfxsize=4.2cm  
\epsffile{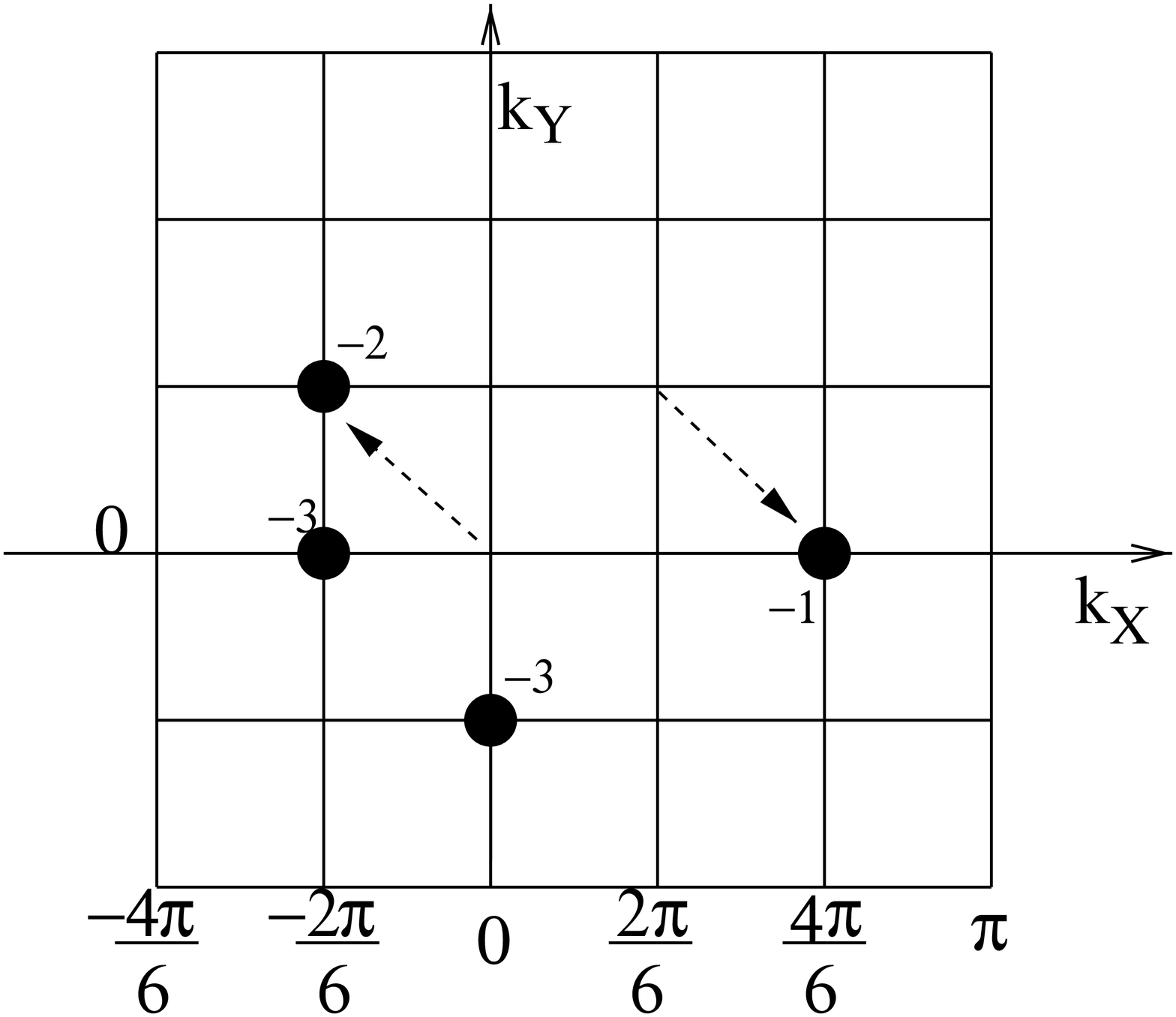}
} 
\caption
{Low kinetic energy plane wave SDs in momentum space of coordinates 
($k_x,k_y$): 1 GS $\left|K_0(\beta)\right>$ 
with $\vec{K}\neq 0$ (upper left), together with 3 plane wave 
SDs of $\vec{K}=0$ and of low energies: one of the 4 
$\left|K_1(\beta)\right>$ of energy $-12t$ (upper right), 
and two of the 16  $\left|K_4(\gamma)\right>$ 
of energy $-9t$ (lower right and left) directly coupled by the two body 
interaction to the $\left|K_1(\beta)\right>$ shown in the upper right figure.
A circle means that the state $(k_x,k_y)$ is occupied, its energy being  
indicated in units of $t$.
}
\label{fig1}
\end{figure}

 When there is no disorder ($W=0$), $\vec{k} =(k_x,k_y)$ being the one 
particle momentum, it is more convenient to write $H$ using the Fourier 
transforms of the creation and annihilation operators. One has the relations
\begin{equation}\label{fouriercj}
c_{\vec{j}} = \frac{1}{L} \sum_{\vec{k}} d_{\vec{k}} e^{i \vec{k}\cdot 
\vec{j}},
\end{equation}
 and
\begin{equation}
d_{\vec{k}} = \frac{1}{L} \sum_{\vec{j}} c_{\vec{j}} e^{-i\vec{k} \cdot 
\vec {j}}. 
\end{equation}
which yield
\begin{equation}\label{Hk}
H = \sum_{\vec{k}} d_{\vec{k}}^{\dagger} d_{\vec{k}} \, \varepsilon(\vec{k}) 
+\sum_{\vec{q},\vec{k}_1,\vec{k}_2} V(\vec{q}) d_{\vec{k}_2+\vec{q}}^{\dagger}
d_{\vec{k}_1-\vec{q}}^{\dagger} d_{\vec{k}_1} d_{\vec{k}_2}
\end{equation}
where 
\begin{equation}\label{states1p}
\varepsilon (\vec{k}) = -2t \left( \cos k_x + \cos k_y \right)
\end{equation}
and
\begin{equation}
V(\vec{q}) = \frac{U}{2L^2} \sum_{\vec{j}\neq \vec{0}} 
\frac{e^{i \vec{q}\cdot \vec{j}}}{\vec{r_{\vec{j0}}}}
\end{equation}

 In the eigenbasis of the non interacting system (eigenvectors 
$d_{\vec{k}_1}^\dagger d_{\vec{k}_2}^\dagger d_{\vec{k}_3}^\dagger 
d_{\vec{k}_4}^{\dagger} \left|0\right>$, $\left|0\right>$ being the 
vacuum state), the Hamiltonian matrix is block diagonal, each block 
being characterized by the same conserved total momentum 
$\vec{K}=\sum_{i=1}^4 \vec{k}_i$. Only the non interacting 
states having in common two $\vec{k}s$ out of four can be coupled by 
the interaction inside a $\vec{K}$ sub-block. Therefore, each 
$\vec{K}$ sub-block is a sparse matrix which can be exactly diagonalized 
using the Lanczos algorithm.

\subsection{The free Fermi limit}
\label{sec:U=0}

When $U=0$, the states are then $N_H$ plane wave SDs
$d_{\vec{k}_1}^\dagger d_{\vec{k}_2}^\dagger d_{\vec{k}_3}^\dagger 
d_{\vec{k}_4}^{\dagger} \left|0\right>$. $N_H=M!/(N!(M-N)!)=58905$ 
for $M=L^2=36$ and $N=4$.  The low energy levels without interaction 
are by increasing energies: 

\begin{itemize}

\item $4$ GSs  $\left|K_0(\beta)\right>$ of energy $E_0(U=0)=-13t$ and 
of momenta $\vec{K}_0 \neq 0$.

\item $25$ first excitations of energy $E_1(U=0)=-12t$,

\item  $64$ second excitations of energy $E_2(U=0)=-11t$ and 
of momenta $\vec{K}_2 \neq 0$. 

\item $180$ third excitations of energy $E_3(U=0)=-10t$ and of momenta 
$\vec{K}_3 \neq 0$.

\item $384$ fourth excitations of energy  $E_4(U=0)=-9t$. 

\end{itemize} 

If one considers the low energy states of total momentum $\vec{K}=0$, 
some of them being shown in Fig. \ref{fig1}, one finds by increasing energy: 

\begin{itemize}

\item  $4$ SDs $\left|K_1(\beta)\right>$ ($\beta=1,\ldots,4$) of energy 
$-12 t$, corresponding to a particle at an energy $-4t$ with 
$\vec{k}_1=(0,0)$, two particles at an energy $-3t$ and a fourth 
particle of energy $-2t$; plus a single SD $\left|K_1(0)\right>$ 
with 4 particles of energy $-3t$.

\item $16$ SDs $\left|K_4(\gamma)\right>$ ($\gamma=1,\ldots,16$) of 
energy $-9t$ given by $8$ SDs where the particles have energies 
$-4t,-3t,-2t,0t$ respectively and by $8$ other SDs where the particles 
have energies $-3t,-3t,-2t,-t$ respectively. Note that the $\left|
K_4(\gamma)\right>$ are directly coupled to the $\left|K_1(\beta)\right>$ 
by the pairwise interaction. 

\end{itemize}

\subsection{The correlated lattice limit}
\label{sec:t=0}

\begin{figure}
\centerline
{ 
\epsfxsize=4.2cm 
\epsffile{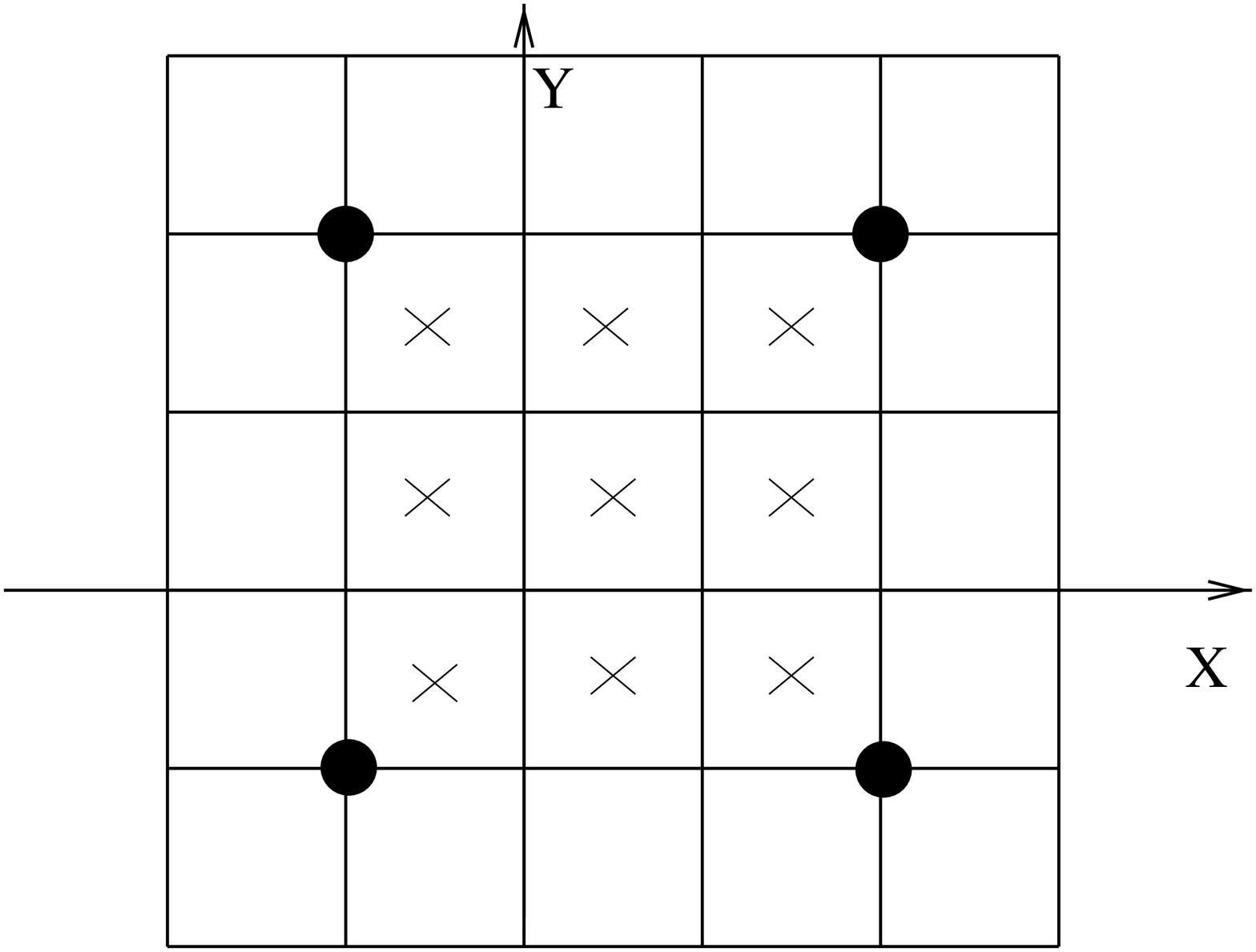}
\hfill 
\epsfxsize=4.2cm  
\epsffile{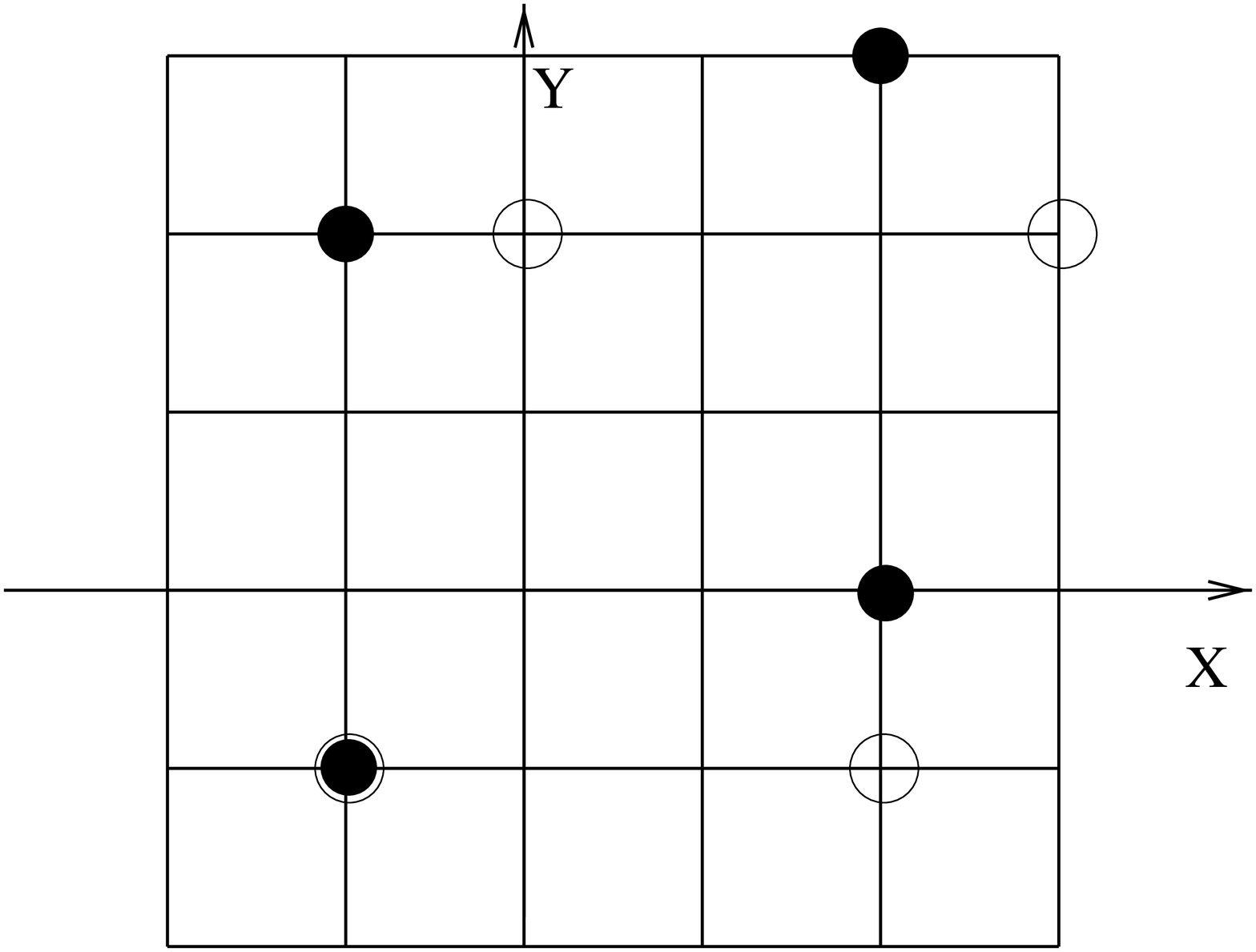}
}
\centerline
{
\epsfxsize=4.2cm  
\epsffile{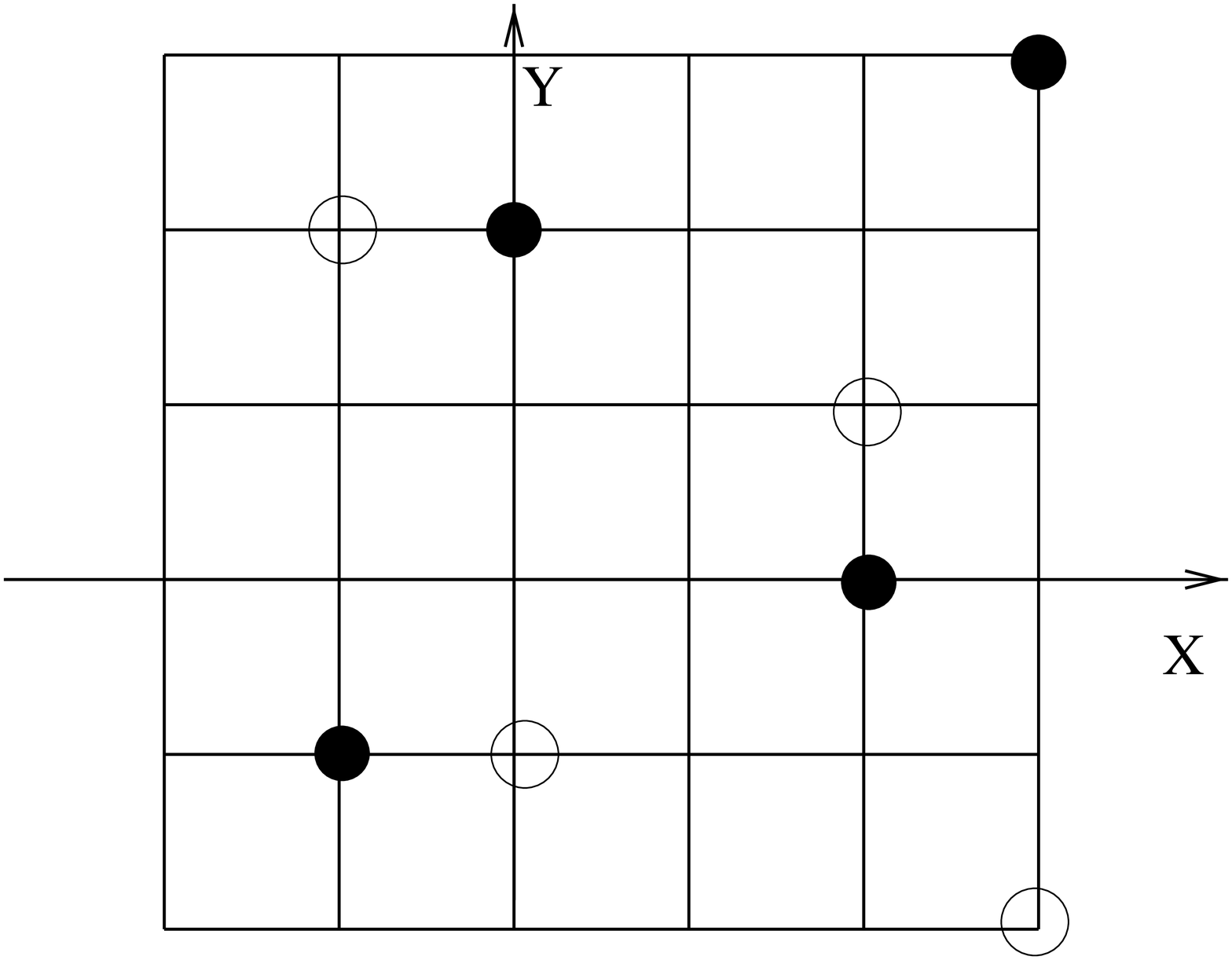}
\hfill 
\epsfxsize=4.2cm  
\epsffile{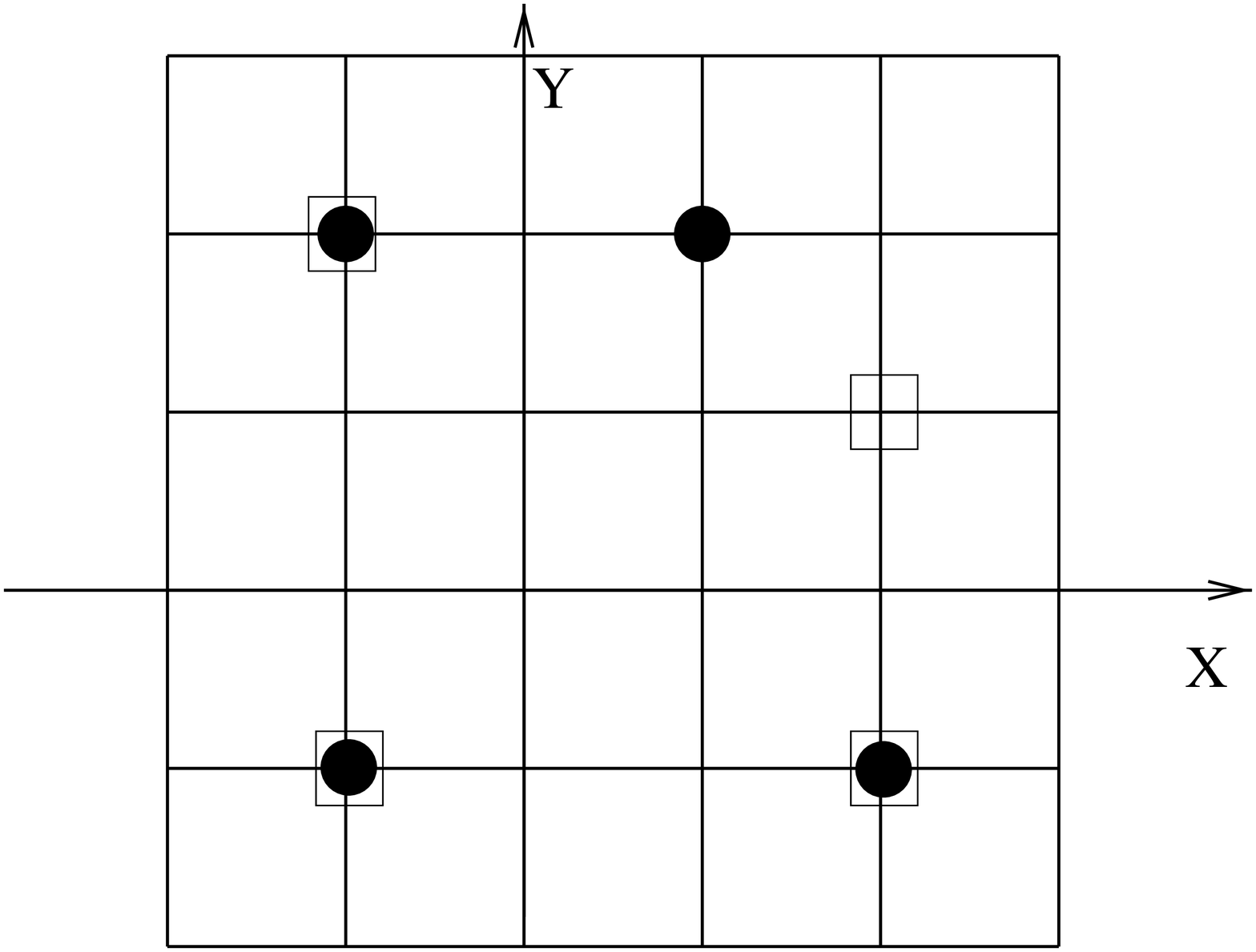} 
}

\caption
{
Low Coulomb energy site SDs in real space of coordinates $(x,y)$: 
one $\left|S\right>$ with 
its 9 possible centers of mass (upper left) , two  $\left|P_2\right>$ 
(upper right), two $\left|P_3\right>$ (lower left) and two $\left|DS\right>$ 
(lower right), which give by successive translations 
one $\left|S(\vec{K}=0)\right>$, two $\left|P_2(\vec{K}=0,J)\right>$, two 
$\left|P_3(\vec{K}=0,J)\right>$ and two out of four 
$\left|DS(\vec{K}=0,J)\right>$ respectively. The two others 
$\left|DS(\vec{K}=0,J)\right>$ are obtained by successive translations 
of the square deformed at the opposite corner.
}
\label{fig2}
\end{figure}

When $t=0$, the translational invariance is broken and the states are 
$N_H$ Slater determinants $c^{\dagger}_{\vec{i}}c^{\dagger}_{\vec{j}}
c^{\dagger}_{\vec{k}}c^{\dagger}_{\vec{l}} \left|0\right>$ built out 
from the site orbitals. The configurations 
$\vec{ijkl}$ correspond to the $N_H$ different patterns characterizing $4$ 
different sites of the $L \times L$ square lattice. The 
configurations of low electrostatic energy are respectively:

\begin{itemize}

\item 9 square configurations $\left|S\right>$ of side 
$b=3$ and of energy $E_0(t=0) \approx 1.80 U$,

\item 36 parallelograms $\left|P_1\right>$ of sides ($3,\sqrt{10}$) 
and of energy $\approx 1.85 U$,

\item 36 other parallelograms $\left|P_2\right>$ of sides 
($\sqrt{10}, \sqrt{10})$ and of energy $\approx 1.97 U$, 

\item $144$ deformed squares $\left|DS\right>$ obtained by moving a single 
site of a square $\left|S\right>$ by one lattice spacing and of energy 
$\approx 2 U$.

\end{itemize}

 Some of those low energy site SDs are shown in Fig. \ref{fig2}. 

 When an infinitesimal hopping term $t$ is included, one 
must delocalize the site SDs in order to restore translational 
invariance and to have eigenstates of given quantized total 
momenta $\vec{K}$. For instance, the $9$ squares $\left|S\right>$ 
give $9$ eigenstates of momentum $\vec{K}$
\begin{equation}
\left|S(\vec{K})\right>=\frac{1}{L^2} \sum_{j_x,j_y=1}^L 
\exp i(\vec{K}\cdot\vec{j}) T_{\vec j} \left|S\right>
\end{equation}
where
\begin{equation}
T_{\vec j} \left|S\right>= 
c^{\dagger}_{(j_x,j_y)} c^{\dagger}_{(j_x,j_y+3)} 
c^{\dagger}_{(j_x+3,j_y)} c^{\dagger}_{(j_x+3,j_y+3)} \left|0\right>.
\end{equation} 
The possible momenta for the $\left|S(\vec{K})\right>$ are given by 
$(K_x,K_y)=2\pi (p_x,p_y) /(L/2)$. 

If we consider the low energy states of total momentum $\vec{K}=0$ 
when $r_s \rightarrow \infty$, one finds by increasing energy the 
following delocalized site SDs: 

\begin{itemize}
 
\item 1 delocalized square $\left|S(\vec{K}=0)\right>$,

\item 2 delocalized parallelograms $\left|P_1(\vec{K}=0,J)\right>$ ($J=1,2$) 
obtained from the 36 $\left|P_1\right>$,

\item 2 delocalized parallelograms $\left|P_2(\vec{K}=0,J)\right>$ ($J=1,2$)
obtained from the 36 $\left|P_2\right>$,

\item 4 delocalized deformed squares $\left|DS(\vec{K}=0,J)\right>$ obtained 
from the 144  $\left|DS\right>$. 

\end{itemize}

\subsection{Level crossing at $r_s^F$ and charge crystallization at $r_s^W$}
\label{sec:crossing}

 The low energy part of the spectrum is shown in Fig. \ref{fig1} 
as a function of $r_s$. If we follow the $4$ GSs of energy $-13t$ at 
$r_s=0$ ($\vec{K}_0\neq 0$), one can see a first level crossing at 
$r_s^F \approx 10$ with a non degenerate level ($\vec{K}_0=0$) which becomes 
the GS above $r_s^F$, followed by two other crossings with two other 
sets of $4$ levels with $\vec{K}_I \neq 0$. When $r_s$ is large, $9$ levels 
coming from $E_1(r_s=0)$ have a smaller energy than the $4$ levels coming 
from $E_0(r_s=0)$.  Since the degeneracies are $(9,36,36)$ when $t=0$, 
these $9$ states give the $9$ square molecules $|S>$ when $r_s 
\rightarrow \infty$. The degeneracies ordered by increasing energy become 
$(1,4,4,4)$ instead of $(4,25,64)$ for $r_s=0$. 

\begin{figure}
\centerline{
\epsfxsize=9cm 
\epsffile{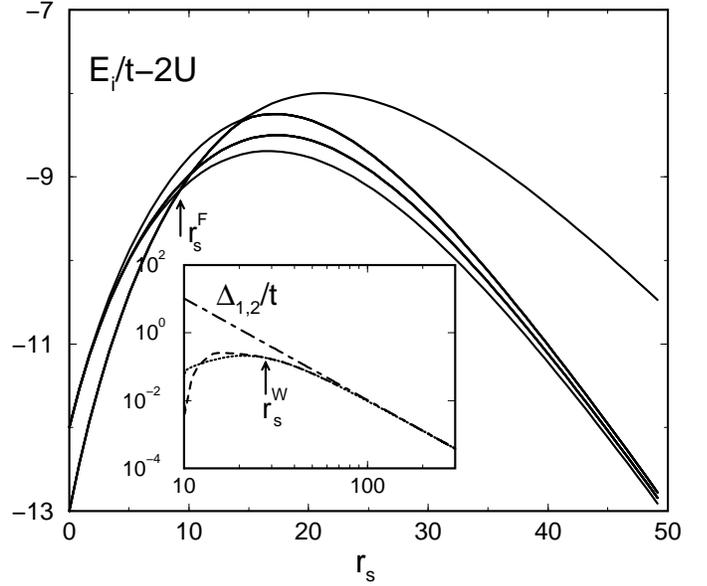}
}
\caption
{As a function of $r_s$, low energy part of the spectrum 
exhibiting a level crossing at $r_s^F$. Inset: two first level spacings 
$\Delta_1/t$ (dashed) and $\Delta_2/t$ (dotted) and the perturbative result 
$3D r_s^{-3}$ (dot-dashed).
}
\label{fig3} 
\end{figure} 

 To describe large $r_s$, one can use degenerate perturbation theory 
and study how the degeneracy of the $9$ $\left|S\right>$ is removed by 
terms $\propto t/U \propto r_s^{-1}$. The centers of mass $\vec{R}$ of the 
9 $\left|S\right>$ are located on the periodic $3 \times 3$ square lattice 
sketched in Fig. \ref{fig2}. For large $r_s$, one has a single 
rigid molecule free to move on this restricted lattice, with a hopping 
term $T\propto t r_s^{-3}$ and a quantized  momentum $\vec{K}=2\pi/(L/2) 
(p_x,p_y)$. Taking into 
account also the corrections to the diagonal matrix elements, one obtains 
for the $9$ first energies $E_{0}({\vec{K}})$ in the limit 
$r_s \rightarrow \infty$ 
\begin{equation}
\frac {E_{0}({\vec{K}})}{t} = 
E_{D} -2 \frac {T}{t} (\cos K_x(I) + \cos K_y(I)).  
\label{eqDPT}
\end{equation}
 $E_{D} = A r_s + B/r_s + C/r_s^3$ and $T/t = D/r_s^3$ 
($A\approx 2.13$, $B\approx -70.81$, $C \approx -18763$ and 
$D \approx 3464$). $E_{D}$ comes from the small vibrations of 
the rigid  molecule while $8T$ is the band width of its zero point 
fluctuations. The degeneracies are $1,4,4$ respectively. 
 
 Four observations can be drawn from this $t/U$ expansion. 
\begin{itemize}

\item The ground state must exhibit a level crossing since 
the total momentum $\vec{K}=0$ when $r_s \rightarrow \infty$  
(lowest quantized kinetic energy for the center of mass of a rigid 
square molecule) while $\vec{K} \neq 0$ when $r_s \rightarrow 0$ 
(incomplete filling of the Fermi shell $-3t$). Is this GS level 
crossing a general feature?  As explained in Ref. \cite{nemeth} 
${\vec K} \neq 0$ at large $r_s$ for $N=3$ and $L=6$, and there 
is no GS level crossing while there is one if $N=3$ 
and $L=8$. When the spins are included, the GS level crossing 
disappears for $N=4$ and $L=6$. $\vec{K}=0$ at large $r_s$ for any 
$L \times L$ square lattice with a filling factor $\nu=1/9$. 
For $L=N=9$, the Fermi shell $-2t$ is totally filled, 
$\vec{K}=0$ at $r_s=0$ and momentum conservation does not yield a 
GS level crossing, in contrast to the case $L=12$ and $N=16$ where the 
Fermi shell is incompletely filled. As we see, the existence of a 
GS level crossing depends on $L$ and $N$ 
and may not have a particular significance. In this work, we have 
studied the true GS, taking the subspaces of $\vec{K} \neq 0$ below 
$r_s^F$, the subspace of $\vec{K}=0$ above $r_s^F$. One could have 
preferred to study the GS inside the subspace of $\vec{K}=0$ 
for all the values of $r_s$, to find that the onset of correlation 
effects which we observe at $r_s^F$ as we will see later should occur 
at a possibly smaller threshold in the $\vec{K}=0$ subspace. 

\item  In the inset of Fig. \ref{fig3}, one can see that the  
$t/U$ expansion gives an accurate description of the $9$ first energies 
above a relatively large value $r_s \approx 100$. As explained in 
Ref. \cite{moises}, this $t/U$ expansion is characteristic of a correlated 
lattice regime where the fluctuations of the charges around the equilibrium 
Wigner lattice sites are strongly restricted by the imposed lattice. This 
lattice expansion has to be distinguished from the large $r_s$ expansion 
of a continuous model, where the oscillatory motion of the electrons around 
the Wigner crystal equilibrium positions gives \cite{wigner} for the GS 
energy an expansion in powers of $1/\sqrt{r_s}$. 

\item Though the $t/U$ lattice expansion ceases to be accurate below 
$r_s \approx 100$, the $9$ low energy states begin to have the structure 
of the spectrum of a single massive particle in a $3 \times 3$ lattice 
(two equal energy spacings $\Delta_1=\Delta_2$ characterizing the 3 first 
sets of states with degeneracies $1,4,4$ respectively) at a lower 
value $r_s^W \approx 28$. This structure means that the system remains 
essentially a rigid square molecule with its $9$ quantized modes for the 
motion of the center of mass down to $r_s^W$. To create a defect in this 
square molecule costs a high energy available in the $10 ^{th}$ excitation 
only. This is why we identify $r_s^W$ as the threshold value for the 
mesoscopic crystallization, above a first threshold $r_s^F$ and below a 
higher threshold $r_s \approx 100$ where the lattice $t/U$ expansion becomes 
valid. 

\end{itemize}

\subsection{Truncated site basis}
\label{sec:truncated}
\begin{figure}
\centerline{
\epsfxsize=8cm 
\epsffile{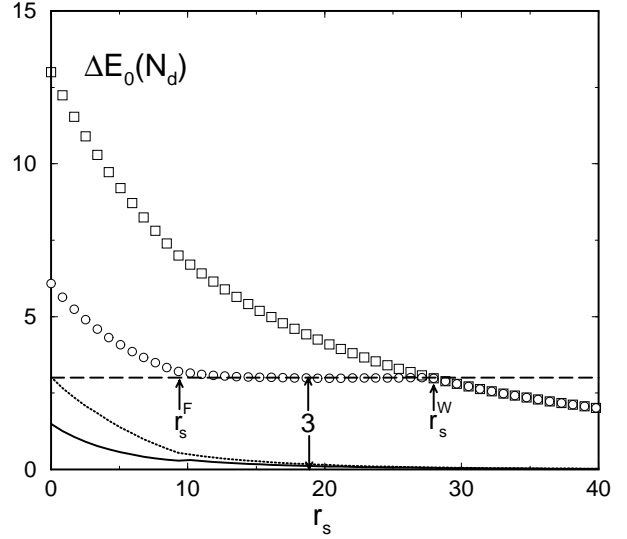}
}
\caption{ 
Errors $\Delta E_0 (N_{d}) = (E_0 (N_d)-E_0)/t$ as a 
function of $r_s$: $d=1$ (thick line), $d=\sqrt2$ (dotted line), 
$d=2$ ($\circ$), $d=\sqrt 5$ ($\square$). 
} 
\label{fig4} 
\end{figure} 

 The $N_H$ site SDs $c^{\dagger}_{\vec{i}}c^{\dagger}_{\vec{j}}
c^{\dagger}_{\vec{k}}c^{\dagger}_{\vec{l}} \left|0\right>$ correspond to the 
$N_H$ different patterns characterizing $4$ 
different sites $\vec{ijkl}$ of the $L \times L$ square lattice. If we order 
those configurations by the smallest distance $d$ between two sites, 
$N_d$ denoting the number of configurations with inter-site spacings 
larger than $d$, one has $N_1=27225, N_{\sqrt2}=9837, N_2=2709, 
N_{\sqrt5}=81$ configurations having a smallest inter-site spacing $>d$, 
out of $N_H=58905$ configurations. The two thresholds $r_s^F$ 
and $r_s^W$ can be also detected if one calculates the GS energy 
$E_0 (N_{d})$ of the truncated Hamiltonian written using the site SDs 
basis restricted to $N_{d}$ site SDs and if we consider the error 
$\Delta E_0 (N_{d}) = (E_0 (N_d)-E_0)/t$ made using this truncation 
for having the exact GS energy $E_0$. As shown in Fig. \ref{fig4}, 
the error $\Delta E_0 (N_1)$ becomes small above $r_s^F$, while 
the error $\Delta E_0(N_{\sqrt 5})=E_0(t=0) - E_0$ for all values of $r_s$. 
The error $\Delta E_0 (N_2)$ has a  very interesting behavior. 
As $r_s$ increases, $\Delta E_0(N_2)$ first decreases up 
to $r_s \approx r_s^F$, then exhibits a very remarkable plateau for 
$r_s^F < r_s < r_s^W$, taking a value  $\approx 3t$ independently of 
$r_s$, before decreasing as $\Delta E_0(N_{\sqrt 5})$ above $r_s^W$. 
This plateau suggests  that the GS for intermediate $r_s$ is 
composed of a floppy  
molecule which can be projected onto the $N_2$ site SDs adapted to describe 
it, plus an unpaired fermion of kinetic energy $\approx -3t$ which is not 
included in this truncated subspace since it is delocalized. Very remarkably, 
this energy turns out to be the energy of a particle at the Fermi surface 
of the non interacting system.

\subsection{GS projections onto plane waves}
\label{k-SDs}

\begin{figure}
\centerline{
\epsfxsize=9cm 
\epsffile{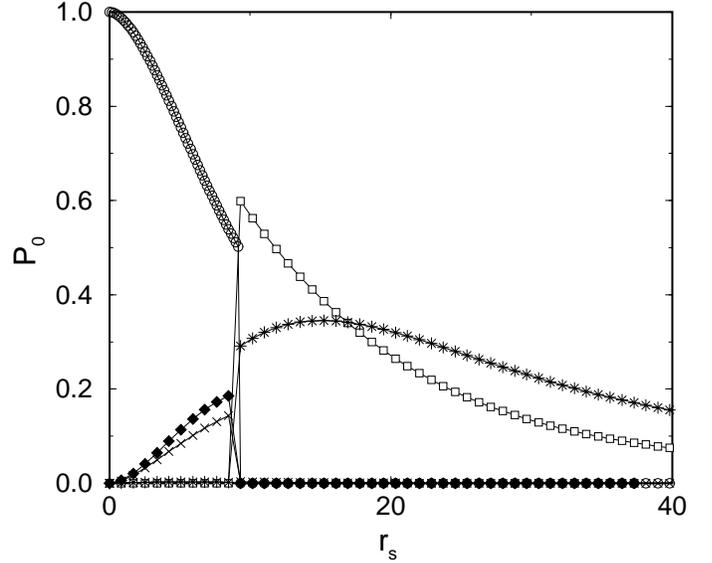}
}
\caption{GS Projections $P_0^0$ ($\circ$), $P_0^1$ ($\square$), 
$P_0^2$ ($\blacklozenge$), $P_0^3$ ($\times$) and $P_0^4$ ($\ast$) 
onto plane wave SDs of low energy when $r_s \rightarrow 0$}
\label{figNFIG1} 
\end{figure} 

 To understand further the nature of the GS between $r_s^F$ and $r_s^W$, 
we have projected the GS wave functions $\left|\Psi_0(r_s)\right>$ over 
the low energy plane wave SDs appropriate to describe unpaired fermions. 
As shown in Fig. \ref{figNFIG1}, below $r_s^F$, a $\vec{K}\neq 0$ GS has a 
large projection $P_0^0(r_s)$ over the $U=0$ GS of same $\vec{K}$ and begins 
to have a smaller projection $P_0^2(r_s)$ over the second excitations of the 
non interacting system of same $\vec{K}$. Above $r_s^F$, the non degenerate 
GS with $\vec{K}=0$ has of course no projection onto the plane wave SDs of 
$\vec{K}\neq 0$, but has a large projection 
\begin{equation}
P_0^1(r_s)=\sum_{\beta=1}^4 |\left<\Psi_0(r_s)| K_1({\beta})\right>|^2 
\end{equation}
which is equally distributed over the $4$ low energy states 
$\left|K_1(\beta)\right>$ of $\vec{K}=0$ and a non negligible projection
\begin{equation}
P_0^4(r_s)=\sum_{\gamma=1}^{16} |\left<\Psi_0(r_s)| K_4({\gamma})\right>|^2 
\end{equation}
over the $16$ previously defined $\left|K_4(\gamma)\right>$ of 
$\vec{K}=0$ which are directly coupled by the interaction to the 
$\left|K_1(\beta)\right>$. 
 
One concludes that a significant part of the system remains an excited 
liquid above $r_s^F$, described by a large projection $P_0=P_0^1+P_0^4$ 
over a few combinations of low energy unpaired fermions. Due to the GS level 
crossing, the intermediate GS has to be described from the $\vec{K}=0$ 
Fermi sea and not from the $\vec{K}\neq 0$ Fermi sea. Since this projection 
is only partial, only a part of the system is made of unpaired fermions, in 
agreement with the concept proposed by Bouchaud et al of a {\it reduced} 
Fermi energy, which decreases as $r_s$ increases.  

\subsection{GS projections onto site orbitals}
\label{site-SDs}

\begin{figure}
\centerline{
\epsfxsize=9cm 
\epsffile{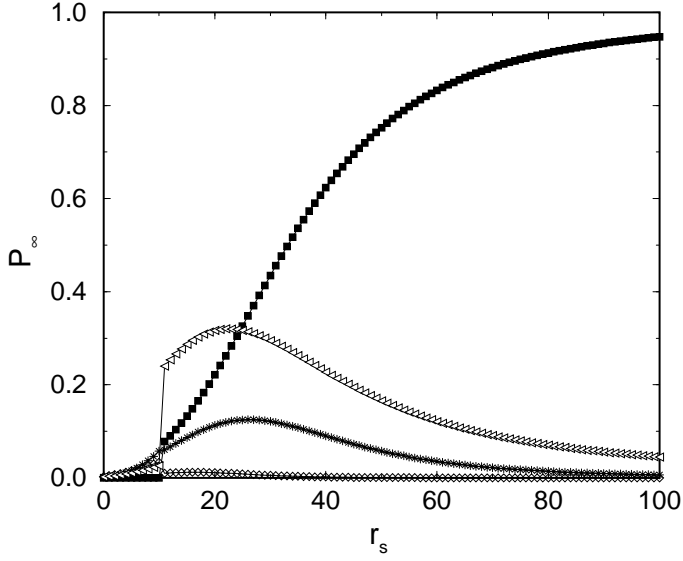}
}
\caption
{GS projection $P_{\infty}^0(r_s)$ ($\blacksquare$), $P_{\infty}^1(r_s)$ 
($\ast$), $P_{\infty}^2(r_s)$ ($\lozenge$) and $P_{\infty}^3(r_s)$ 
($\lhd$) onto the first $\vec{K}=0$ delocalized site SDs 
of low energy when $r_s \rightarrow \infty$. 
}
\label{NFIG2}  
\end{figure} 

 We study now the GS projection over the low energy site orbitals 
shown in Fig. \ref{fig2}, which become the eigenstates when 
$r_s \rightarrow \infty$. More precisely, we consider 
the first delocalized site SDs having a delocalized center of mass and 
the same momentum $\vec{K}$ than the GS: The delocalized square 
$\left|S(\vec{K})\right>$, the 2 delocalized parallelograms 
$\left|P_1(\vec{K},J)\right>$, the 2 delocalized 
parallelograms $\left|P_2(\vec{K},J)\right>$, and the 4 delocalized 
deformed squares $\left|DS (\vec{K},J)\right>$. 
Fig. \ref{NFIG2} shows the behaviors of the GS projections 
\begin{equation}
P_{\infty}^0(r_s)=|\left<\Psi_0(r_s)|S(\vec{K})\right>|^2 
\end{equation}
\begin{equation}
P_{\infty}^1(r_s)= \sum_{J=1}^2|\left<\Psi_0(r_s)|P_1(\vec{K},J)\right>|^2 
\end{equation}
\begin{equation}
P_{\infty}^2(r_s)= \sum_{J=1}^2|\left<\Psi_0(r_s)|P_2(\vec{K},J)\right>|^2 
\end{equation}
\begin{equation}
P_{\infty}^3(r_s)=|\left<\Psi_0(r_s)|DS (\vec{K},J)\right>|^2,  
\end{equation}
where $\vec{K}$ is the GS momentum ($\vec{K}\neq 0$ below $r_s^F$ and 
$\vec{K}=0$ above $r_s^F$). 

 While a $\vec{K}\neq 0$ GS has negligible projections 
over the low energy site SDs of same $\vec{K}$ below $r_s^F$, 
there is an important contribution above $r_s^F$ of the deformed 
squares, of the square and of the parallelograms 1 of $\vec{K}=0$. 
As $r_s$ increases, the GS projection $P_{\infty}^0(r_s)$ over the square 
molecule of $\vec{K}=0$ goes to 1. The GS  projection $P_{\infty}^3(r_s)$  
over the $4$ deformed squares of $\vec{K}=0$ is the main projection 
below $r_s \approx r_s^W$, a threshold value where the GS projection 
$P_{\infty}^1(r_s)$ over the parallelograms 1 is maximum. 
One concludes that above $r_s^F$, the missing part of the system, 
which is not described by the low energy unpaired fermions of 
the previous section, is a floppy Wigner molecule, mainly made of 
deformed squares below $r_s^W$ and of squares above $r_s^W$.

\subsection{GS projections onto a combined basis of plane waves 
and site orbitals}
\label{sec:combined}

\begin{figure}
\centerline{
\epsfxsize=9cm 
\epsffile{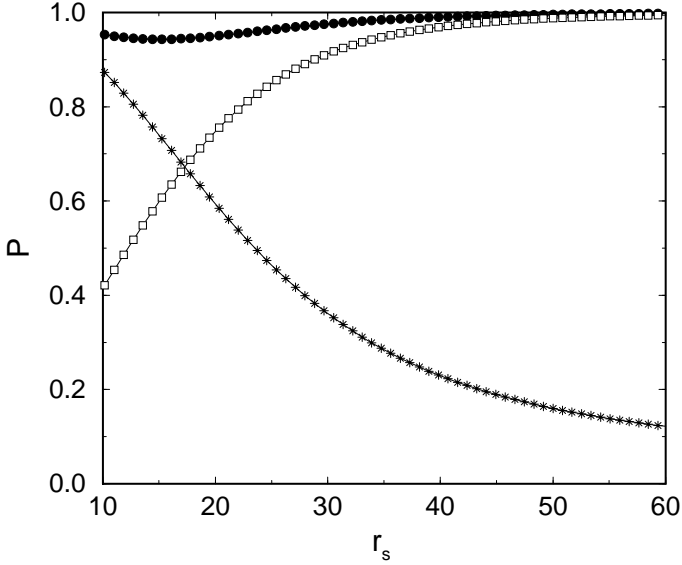}
}
\caption
{GS projection $P_0(r_s)$ ($\ast$) over the $20$ $\vec{K}=0$ plane wave SDs 
of low energy, $P_{\infty}(r_s)$ ($\square$) over the $9$ $\vec{K}=0$ 
delocalized site SDs and $P_t(r_s)$ ($\bullet$) over combined plane wave and 
site SD re-orthonormalized basis.
}
\label{NFIG3}  
\end{figure} 

The site SDs and plane wave SDs are not orthonormal. After 
re-orthonormalization, the total GS projection $P_t(r_s)$ 
over the subspace spanned by the $20$  plane wave SDs 
and the $9$  delocalized site SDs of low energy and of 
momentum $\vec{K}=0$ is given in Fig. \ref{NFIG3}, together with 
the GS projection $P_0(r_s)=P_0^1(r_s)+P_0^4(r_s)$ over the 
$20$ plane wave SDs and $P_{\infty}(r_s) =\sum_{I=0}^3 
P_{\infty}^I(r_s) $ over the $9$ delocalized site SDs of 
momentum $\vec{K}=0$. One can see than more than $95/100$ of 
the intermediate GS is located inside this combined subspace, 
suitable to describe a floppy solid co-existing with low energy 
unpaired fermions. This demonstrates the Andreev-Lifshitz conjecture 
for the considered mesoscopic lattice model. 

 Let us point out that those exact results raise serious objections about 
the validity of ``exact'' studies of a few particles in the continuum, 
where the infinite Hilbert space is truncated to the finite basis made 
of the low energy plane wave SDs only. For intermediate $r_s$, our exact 
results show the necessity to combine plane wave SDs and site 
SDs before truncating. 

\subsection{Inter-particle spacings}
\label{sec:Spacings}
 
\begin{figure}
\centerline{
\epsfxsize=8cm 
\epsffile{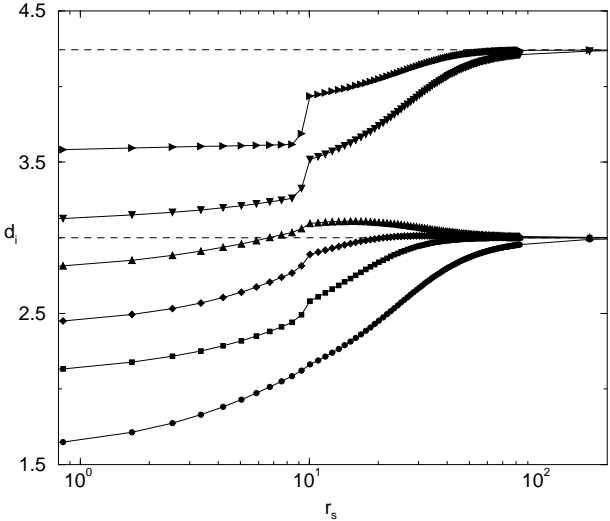}
}
\caption{ 
Disorder average of the $6$ mean inter-particle spacings $\left<d(p)\right>$ 
as a function of $r_s$.
} 
\label{fig11} 
\end{figure} 

 To understand the nature of the intermediate GS, we study the 
distribution of the different inter-particle spacings. 
For the site SDs  $c^{\dagger}_{\vec{i}}c^{\dagger}_{\vec{j}}
c^{\dagger}_{\vec{k}}c^{\dagger}_{\vec{l}} \left|0\right>$, 
one defines the $6$ spacings $d_{\vec{ijkl}}(1) 
\leq d_{\vec{ijkl}}(2) \leq \ldots \leq d_{\vec{ijkl}}(6)$ 
of each configuration $\vec{ijkl}$ ordered by increasing values. The 
$n^{th}$ moment $d^n(p)$ of the $p^{th}$ GS inter-particle 
spacing  at $r_s$ is given by: 
\begin{equation}
\left<d^n(p)\right> =\sum_{\vec{ijkl}=1}^{N_H} d_{\vec{ijkl}}^n(p) 
|\left<\Psi_0(r_s)\right|c^{\dagger}_{\vec{i}}c^{\dagger}_{\vec{j}}
c^{\dagger}_{\vec{k}}c^{\dagger}_{\vec{l}} \left|0\right>|^2. 
\end{equation}
Very weak random potentials are included to get rid of the symmetries of 
the $6 \times 6$ lattice. After average over an ensemble of random 
configurations, we show in Fig. \ref{fig11} how the $6$ mean 
GS inter-particle spacings $\left<d_p\right>$ vary as a function of $r_s$ 
for a value $W=0.1$ of the disorder strength. When 
$r_s \rightarrow \infty$, the $3 \times 3$  Wigner molecule 
gives $d_1=d_2=d_3=d_4=3$ and $d_5=d_6=3 \sqrt{2}$. As $r_s$ decays, one can 
see that one of the largest spacings out of two and two of the smallest 
spacings out of four remain close their asymptotic values, in contrast to 
the others. This shows us that one has for intermediate $r_s$ a 
floppy solid made of three particles, while the fourth particle 
remains delocalized. A similar conclusion was drawn  from a study of 
the case $N=3$ and $L=6$ in Ref. \cite{nemeth}, 
where the intermediate GS was shown to be a floppy two particle molecule 
co-existing with a third delocalized particle. 
\begin{figure}
\centerline{
\epsfxsize=8cm 
\epsffile{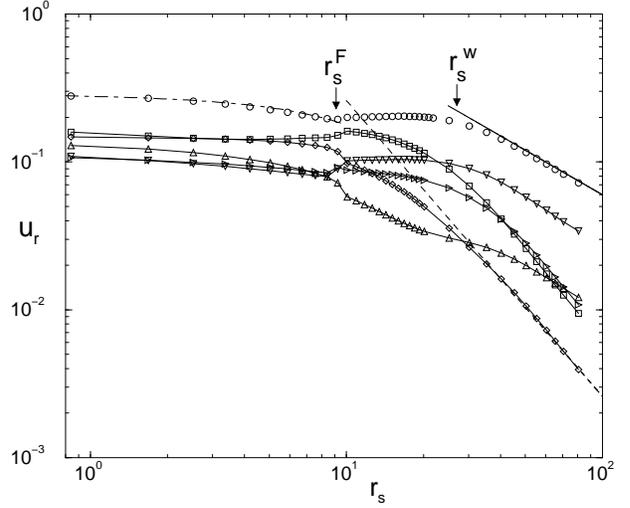}
}
\caption{ 
Disorder average of the relative fluctuations $\left<u_d (i)\right>$ 
of the $6$ inter-particle spacings as a function of $r_s$, 
using the same symbols as in Fig. \ref{fig11}. Characteristic behaviors 
$u_r=0.28 - 0.01 r_s$ (dotted-dashed), $u_r \propto 6/r_s$ (solid) and 
$u_r \propto 26/r_s^2$ (dashed). Note the three spacings having an almost 
$r_s$ independent fluctuations for $ r_s^F < r_s < r_s^W$.  
} 
\label{fig12} 
\end{figure} 

 The behaviors of the relative fluctuations 
\begin{equation}
u_d(p)=\sqrt{\frac{\left<d^2(p)\right>}{\left<d(p)\right>^2}-1}
\end{equation}
of the $6$ inter-particle spacing $d(p)$ are given in Fig. \ref{fig12}.
When $r_s \rightarrow \infty$, the fluctuations of the square molecule 
can be calculated using the $t/U$ lattice expansion. At first order, one 
can move only a single particle, which modifies three inter-particle 
spacings out of six. The fluctuations of the three remaining spacings is 
obtained by moving two particles, which requires to go to the second order. 
This explains the three $r_s^{-1}$ decays and the three $r_s^{-2}$ decays 
characterizing the correlated lattice regime. The behaviors in the 
intermediate regime are remarkable: 

\begin{itemize}

\item The relative fluctuations of three spacings out 
of six decay as $r_s$ increases, as one can expect if a floppy 
$3$ particle molecule becomes more rigid as $r_s$ increases.

\item The three others spacings have relative fluctuations 
which are nearly independent of $r_s$ between $r_s^F$ and $r_s^W$, as 
one can expect if the $4^{th}$ particle remains delocalized. 

\item Notably, the smallest spacing has a relative fluctuation 
which varies as $0.28 - 0.01 r_s$ for $r_s < r_s^F$ (weak coupling 
regime), which is almost independent of $r_s$ for $r_s^F < r_s < r_s^W$ 
(intermediate regime) before decreasing as $6/r_s$ above $r_s^W$ 
(correlated lattice regime).  

\end{itemize}

The behaviors of the different inter-particle spacings are consistent with 
the Andreev-Lifshitz conjecture for the considered mesoscopic lattice model. 
Notably, the number of crystal lattice sites is indeed smaller than the 
total number of particles.

\subsection{GS response to small perturbations}
\label{sec:Perturbation}

 We study now the GS response to small perturbations, when the 
site potentials are non random ($W=0$), which gives us  
other signatures of the intermediate regime. 

\subsubsection{Aharonov-Bohm flux}
\label{sec:flux}

\begin{figure}
\centerline{
\epsfxsize=8cm 
\epsffile{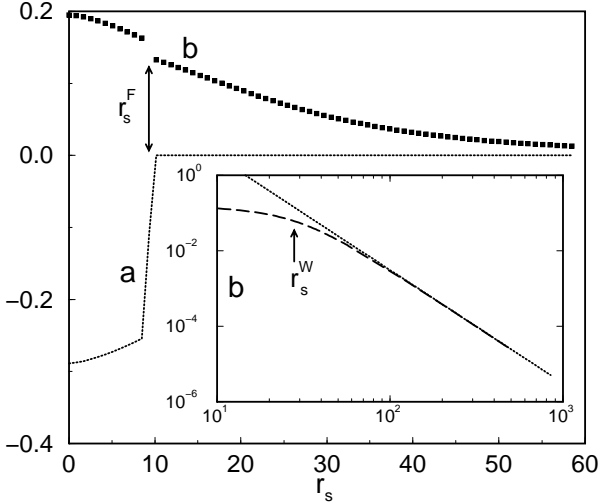}
}
\caption
{ 
Functions $a(r_s)/t$ (dotted) and $b(r_s)/t$ 
(filled squares) characterizing the GS response to an infinitesimal flux; 
inset : $b(r_s)/t$ (dashed) and $8D/9 r_s^{-3}$ (dotted).
}
\label{FLUX} 
\end{figure} 

 The first one consists in piercing the $2d$ torus by an infinitesimal 
positive flux $\phi$ (periodic BCs along the $y$ direction, $t \rightarrow 
t\exp(i \phi /L)$ for hopping along the $x$-direction only, 
$\phi=\pi$ corresponding to anti-periodic BCs).  
The coefficients $a(r_s)$ and $b(r_s)$ (Kohn curvature) of the expansion 
$E_0(r_s,\phi) \approx E_0(r_s,0) + a(r_s) \phi+ b(r_s) \phi^2 /2$ are 
given in Fig. \ref{FLUX}. When $r_s=0$, $\phi$ removes the 
fourfold degeneracy of $E_0$, $a=-\sqrt{3}t/6$ and $b = 7t/36$. 
When $r_s$ is large, the substitution $K_x(I) \rightarrow K_x(I)+2\phi/3$  
in Eq. \ref{eqDPT}  gives $a=0$ and $b \approx 8Dtr_s^{-3}/9$. 
An infinitesimal positive flux $\phi$ gives rise to a persistent current 
$I_x =-\partial E_0 /\partial \phi = - a$ when $r_s < r_s^F$ while the GS 
curvature $b$ exhibits a smooth crossover between two regimes around $r_s^W$ 
(inset of Fig. \ref{FLUX}). 

\subsubsection{Single pinning well}
\label{sec:well}

\begin{figure}
\centerline{
\epsfxsize=8cm 
\epsffile{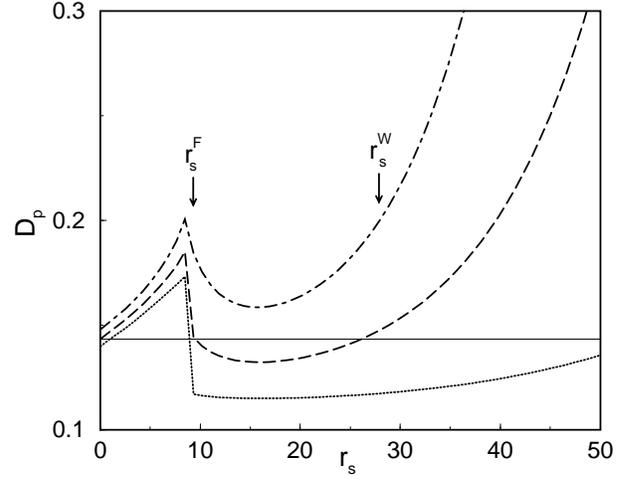}
}
\caption
{
GS density $D_{\vec{p}}(r_s)$ at the pinning site $\vec{p}$ with 
$V_{\vec{p}}/t=-0.01$ (dotted), $-0.05$ (dashed) and $-0.1$ (dot-dashed).  
}
\label{NFIG4} 
\end{figure} 

The second perturbation consists in introducing a weak 
negative potential $V_{\vec{p}}$ at a single lattice site $\vec{p}$. 
The GS density   
\begin{equation}
D_{\vec{p}}(r_s)=\left<\Psi_0(r_s)\right|c^{\dagger}_{\vec{p}} 
c_{\vec{p}} \left|\Psi_0(r_s)\right>
\end{equation}
at the site $\vec{p}$ is shown in Fig. \ref{NFIG4}. If 
$V_{\vec{p}}=0$, $D_{\vec{p}}(r_s=0)=1/9$. A weak negative value of 
$V_{\vec{p}}$ yields a larger value for $D_{\vec{p}}(r_s=0)$. When one turns 
on the interaction, $D_{\vec{p}}$ first increases and drops at $r_s^F$, 
where the interacting GS begins to have a weaker response to a 
weak pinning well than the non interacting GS. When $r_s$ 
is large, $D_{\vec{p}}$ increases again and the Wigner molecule is pinned. 
This surprisingly weak response for intermediate $r_s$ suggests that 
the system may very weakly respond to the presence of weak impurities. 
Let us underline that this is precisely for those values of $r_s$ 
that the new $2d$ metal has been observed in $2d$ field effect 
devices \cite{kravchenko}.

\section{Lattice model with random potentials}
\label{sec:crossover}

 We extend our study of the disorder by increasing $W$ from the 
previously considered weak value ($W=0.1$) up to larger values ($W 
\rightarrow 5$) which are too small for having exponential localization 
of the one particle states on a scale $L=6$, but sufficient for having one 
particle diffusion. 

\subsection{Inter-particle spacings}
\label{sec:spacing-dis}

\begin{figure}
\centerline{
\epsfxsize=8cm 
\epsffile{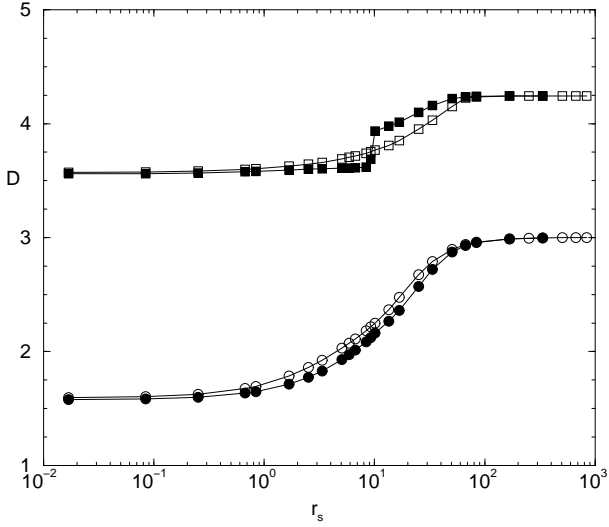}
}
\caption{ Disorder average of the mean smallest and largest 
inter-particle spacings for $W=0.1$ (filled symbols) and $5$ 
(empty symbols) as a function of $r_s$.  
} 
\label{fig13} 
\end{figure}

\begin{figure}
\centerline{
\epsfxsize=8cm 
\epsffile{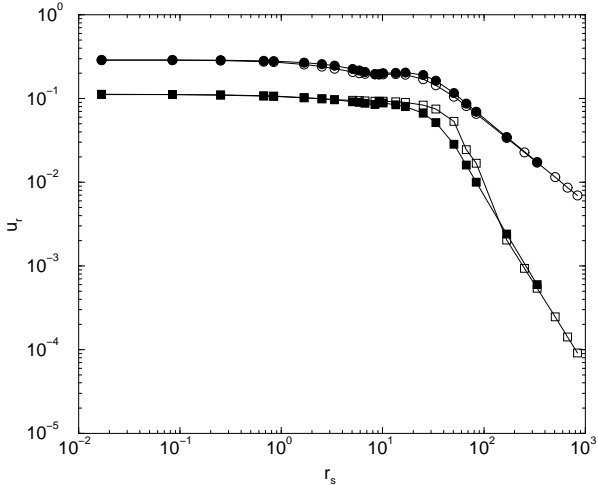}
}
\caption{ 
Disorder average of the relative fluctuations of the smallest and 
largest inter-particle spacings as a function of $r_s$ for 
$W=0.1$ and $5$ (same symbols as in Fig. \ref{fig13}).  
} 
\label{fig14} 
\end{figure} 

 In Fig. \ref{fig13}, one can compare the mean smallest and largest 
inter-particle spacings. For $W=0.1$, the effect of the level crossing 
at $r_s^F$ is still visible when one follows the largest spacing, 
but becomes very smooth for the smallest spacing. The jump associated to 
$r_s^F$ is totally smeared for $W=5$. One can also see that the 
smallest spacing is closer to its asymptotic value $3$ as one increases 
$W$, showing that disorder favors the formation of a correlated 
glass. In contrast, the disorder defavors the formation of a perfect 
crystalline array, the largest spacing  requiring a larger $U$ to 
reach its asymptotic value $3\sqrt{2}$. Similar conclusion can be drawn from  
Fig. \ref{fig14} where the corresponding relative fluctuations of the 
smallest and largest spacings are given.

\subsection{First energy excitation}
\label{sec:first excitation}
 
\begin{figure}
\centerline{
\epsfxsize=8cm 
\epsffile{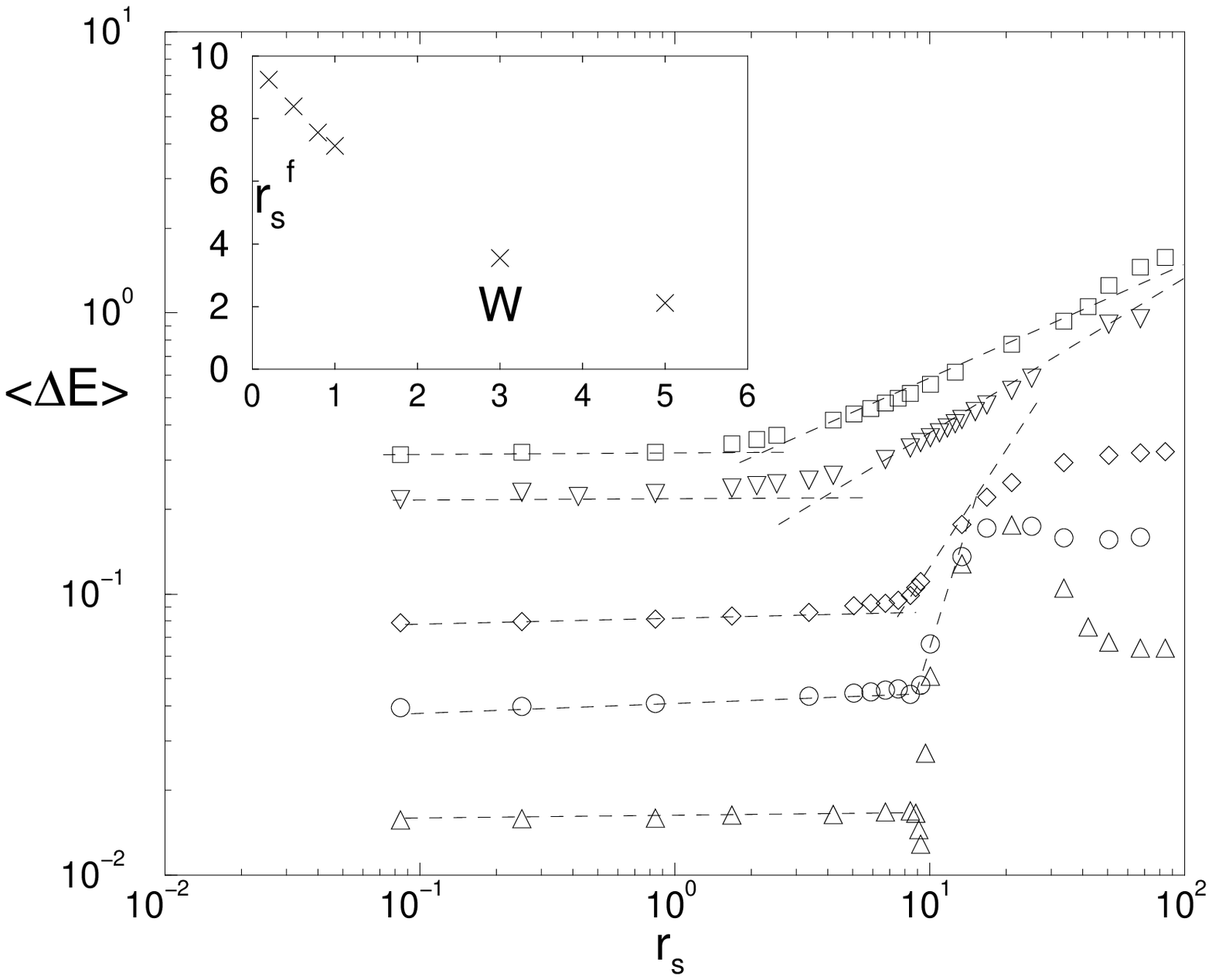}
}
\caption{ 
Ensemble average of the first energy spacing 
$\left<\Delta E\right>=\left<E_1-E_0\right>$ as a function of 
$r_s$ for $W=0.2$ ($\triangle$), $0.5$ ($\circ$), 
$1$ ($\diamond$), $3$ ($\nabla$) and $5$ ($\Box$). Inset: crossover 
values $r_s^F(W)$ as a function of $W$. 
} 
\label{fig15} 
\end{figure} 

 Fig. \ref{fig15} gives the average over the disorder of the first 
energy spacing $\left<\Delta E\right>=\left<E_1-E_0\right>$ as a function 
of $r_s$ for 
different values of $W$. One can see a weak coupling regime where 
$\left<\Delta E\right>$ does not depend on $r_s$, followed by an increase of 
$\left<\Delta E\right>$ as a function of $r_s$. The threshold coupling  
between those two regimes is naturally of the order of $r_s\approx 10$ when  
$W$ is small. When $W$ is larger, $\left<\Delta E\right> 
\propto r_s^{\alpha}$ as indicated by the dashed lines of Fig. \ref{fig15}. 
Assuming that the correlated regime occurs at the ratios $r_s^F(W)$ below 
which $\left<\Delta E\right>$ does not depend on $r_s$, and above which 
$\left<\Delta E\right>$ increases as $r_s^{\alpha}$, we have plotted 
in the inset of Fig. \ref{fig15} how $r_s^F(W)$ depends on $W$. The 
onset $r_s^F(W)$ of the correlation effects decays from the value 
$\approx 10$ where there is a level crossing without disorder towards a 
much smaller value when $W \rightarrow 5$.

\begin{figure}
\centerline
{ 
\epsfxsize=8cm 
\epsffile{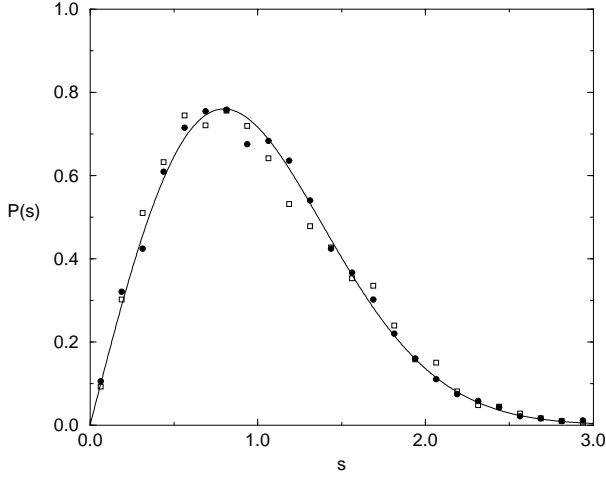}
} 
\caption
{
 Distribution $P(s)$ of the first energy excitation 
$s=E_1-E_0/\left<E_1-E_0\right>$ for $W=5$ and $r_s=1.7$ ($\bullet)$ 
and $r_s=5$ ($\Box$). The solid line is the Wigner surmise $P_W(s)$. 
}
\label{fig16}
\end{figure}

\begin{figure}
\centerline
{ 
\epsfxsize=8cm 
\epsffile{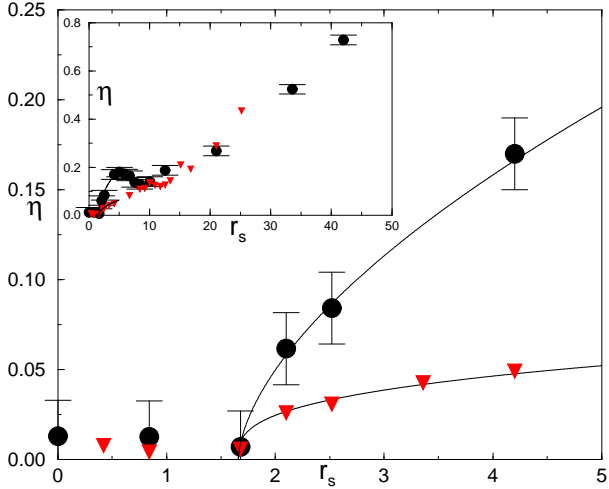}
} 
\caption
{
Spectral parameter $\eta$ for the first energy spacing as a 
function of $r_s$ for $W=3$ ($\blacktriangledown$) and $5$ 
($\bullet$), showing the 
sharp breakdown of the Wigner distribution at $r_s^F \approx 1.7$. 
The inset shows that the level repulsion is totally suppressed 
when $r_s\rightarrow \infty$.
}
\label{fig17}
\end{figure}

This onset $r_s^F (W)$ for $W= 3\ldots 5$ can also be seen in the 
sample-to-sample distribution of the first energy excitation. When 
$r_s$ is weak, the first many body excitation corresponds to a single 
one body excitation above the Fermi energy. If the one body motion is 
diffusive, the one body spectrum is correlated, and the distribution 
$P(s)$ of the energy spacing between consecutive levels is given by 
the Wigner surmise $P_W(s)$. Therefore, the first many body spacing 
$s=E_1-E_0/\left<E_1-E_0\right>$ measured in units of its ensemble 
averaged value is given by the Wigner surmise: 
\begin{equation} 
P_W (s)= \frac{\pi s}{2}\exp(-\frac{\pi s^2}{4})
\end{equation}
as one can see in Fig. \ref{fig16} for $r_s < r_s^F(W)$. 
When $r_s$ exceeds $r_s^F(W)$, the level repulsion becomes weaker. 
As shown in Fig. \ref{fig16}, $P(s)$ for low $s$ 
is systematically larger than $P_W(s)$ when $r_s =5$. To study this 
weakening  of the spectral rigidity, we define a spectral parameter 
\begin{equation}
\eta (r_s)=\frac{\int_0^a (P(s)-P_W(s)) ds}
{\int_0^a (P_P(s)-P_W(s)) ds},
\end{equation} 
where $P_P(s)=\exp -s$ is the Poisson distribution characterizing 
uncorrelated levels and $a=0.4729$ is the value where 
$P_W(a)=P_P(a)$. $\eta=1$ when $P(s)=P_P(s)$ and $\eta=0$ when 
$P(s)=P_W(s)$. Very remarkably, one can see in  Fig. \ref{fig17} 
that the first energy excitation is well described 
by the Wigner surmise up to a threshold consistent with $r_S^F(W)$ 
where the spectral rigidity suddenly becomes weaker. The curves 
$\eta (r_s)$ are given on a larger interval of values of $r_s$ 
in the inset of Fig. \ref{fig17}, and one can see the two 
characteristic thresholds detected in earlier studies \cite{bwp1,bwp2} 
for $W=5$: $r_s^F(W=5)\approx 2$ where one has the breakdown of 
Wigner-Dyson rigidity and $r_s^W(W=5)\approx 10$ where was located the 
onset of charge crystallization. 

In summary, when disorder yields one particle diffusion, the onset 
of correlation effects occurs at a weaker value $r_s^F(W)$ than in 
the clean limit, yielding for the excitation energy $\Delta E$ an 
increase of its average value when $r_s$ increases and a change of 
its distribution. 

\subsection{Breakdown of the Hartree-Fock approximation}
\label{sec:persistent currents}

\begin{figure}[ht]  
\vskip.2in
\centerline{\epsfxsize=8cm\epsffile{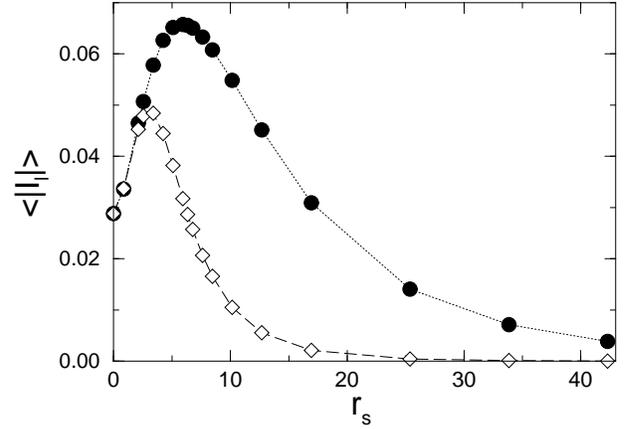}} 
\caption{Ensemble average GS current $\left<I_x\right>$  
as a function of $r_s$ for $W=5$. Exact 
values (filled symbols) and HF values (empty symbols). 
}
\label{fig-HF} 
\end{figure}

 A last evidence proving that $r_s^F(W)$ is indeed the onset of 
the correlation effects is shown by Fig. \ref{fig-HF} where 
the disorder average total GS persistent current $\left<I_x\right>$ 
enclosing a flux $\phi=\pi/2$ is shown as a function of $r_s$. Both 
the exact numerical value and the mean field value given by the Hartree-Fock 
approximation are given. As one can see, the change seen in 
the mean and in the distribution of the first energy excitation at 
$r_s^F(W)$ are correlated with the fact that the ground state 
cannot been described by the best possible Slater 
determinant (HF-approximation), due to correlation 
effects which are beyond a simple mean field approach.

\section{Conclusion}
\label{sec:conclusion}

 Firstly, it is interesting to compare our results obtained on a $2d$ torus 
without edge to those obtained using an harmonic confinement. For 
instance, a Monte Carlo study \cite{filinov} of a few electrons 
in an harmonic trap concludes that mesoscopic Wigner 
crystallization proceeds in two stages: (i) via radial ordering of 
electrons on shells and (ii) freezing of the intershell rotation. 
This crystallization in two steps, with a particular 
intermediate behavior, could be attributed to the non uniform 
density characterizing the harmonic trap. One may argue that 
crystallization takes place in the low density edges before 
the large density bulk, such that this intermediate regime 
might be related to some interplay between edge and bulk 
effects. We find that mesoscopic Wigner crystallization takes also 
place in two stages when the particles are confined on a $2d$ torus 
with a uniform density. 

 Secondly, this raises the question to know if this intermediate regime 
is a pure mesoscopic effect valid only in small systems, or the mesoscopic 
trace of the Andreev-Lifshitz supersolid, where unpaired fermions with 
reduced Fermi energy co-exist with a floppy solid. This might explain 
a few recent experimental studies \cite{gao1,gao2,illani} of the 
metallic $2d$ hole gas in GaAs, where a not well identified metallic 
phase seems to coexist with a more usual Fermi liquid phase, responsible 
of usual weak localization behaviors. This two-phase coexistence scenario 
may be simply explained by a supersolid phase, without having to necessarily 
take into account the disorder effects \cite{Shi}. Upon completion of this 
manuscript, we received from Boris Spivak \cite{spivak} a preprint where 
the existence of an intermediate phase between the Fermi liquid and the 
Wigner crystal is claimed to be a generic property of the $2d$ pure electron 
liquid in MOSFETs at zero temperature, and where the consequences for the 
experimental results obtained in $2d$ MOSFETs are also discussed.  

 Thirdly, one cannot exclude that a supersolid regime is favored 
by the chosen geometry, because of the number of particles and 
underlying square lattice, but not favored at all in the continuous limit, 
which has not a square symmetry but a spontaneously broken hexagonal 
symmetry. 

\begin{acknowledgement}

We thank G. Benenti for the calculation of the H-F persistent 
currents in subsection \ref{sec:persistent currents},  B. Spivak 
and J.-P. Bouchaud for drawing our attention onto Refs. \cite{andreev} and 
\cite{bouchaud} respectively and Z. \'A. N\'emeth for stimulating 
discussions. G. Katomeris acknowledges the financial support provided 
through the European Community's Human Potential Programme under contract 
HPRN-CT-2000-00144.  
 
\end{acknowledgement}

\end{document}